\documentclass[useAMS,usenatbib,onecolumn,usegraphicx]{mn2e}

\newcommand{\beq}{\begin{equation}}
\newcommand{\eeq}{\end{equation}}
\newcommand{\beqar}{\begin{eqnarray}}
\newcommand{\eeqar}{\end{eqnarray}}

\title[Torsional Oscillations of Relativistic Stars] {Torsional Oscillations of Relativistic Stars with Dipole Magnetic Fields}
\author[H.~Sotani, K.~D. Kokkotas and N.~Stergioulas] {H.~Sotani\thanks{E-mail:
sotani@astro.auth.gr}, K.~D.~Kokkotas\thanks{E-mail:
kokkotas@auth.gr} and N.~Stergioulas\thanks{E-mail:
niksterg@astro.auth.gr}
\\
  Department of Physics, Aristotle University of Thessaloniki,
  Thessaloniki 54124, Greece}
\begin{document}



\maketitle

\label{firstpage}

\begin{abstract}
  We present the formalism and numerical results for torsional
  oscillations of relativistic stars endowed with a strong dipole
  magnetic field, assumed to be confined to the crust. In our approach, we focus on
  axisymmetric modes and neglect higher-order couplings induced by the
  magnetic field.  We do a systematic search of parameter space by
  computing torsional mode frequencies for various values of the
  harmonic index $\ell$ and for various overtones, using an extended
  sample of models of compact stars, varying in mass, high-density
  equation of state and crust model.  We show that torsional mode
  frequencies are sensitive to the crust model if the high-density
  equation of state is very stiff. In addition, torsional mode
  frequencies are drastically affected by a dipole magnetic field, if
  the latter has a strength exceeding roughly $10^{15}$G and we find
  that the magnetic field effects are sensitive to the adopted crust
  model. Using our extended numerical results we derive empirical
  relations for the effect of the magnetic field on torsional modes as
  well as for the crust thickness. We compare our numerical results to
  observed frequencies in SGRs and find that certain high-density EoS
  and mass values are favored over others in the non-magnetized
  limit. On the other hand, if the magnetic field is strong, then its
  effect has to be taken into account in attempts to formulate a
  theory of asteroseismology for magnetars.
\end{abstract}

\section{Introduction}
\label{sec:Intro}


There is growing evidence that Soft Gamma Repeaters (SGRs) could be
magnetars experiencing starquakes that are connected (through the
intense magnetic field) to gamma ray flare activity. Magnetars are
thought to be neutron stars with very strong magnetic fields,
greatly exceeding $\sim 5 \times 10^{13}$ G \citep{DT1992}. The
flare activity consist of an initial short peak in the hard part of
the spectrum, followed by a decaying softer part (tail) which last
for hundreds of seconds. Recent analysis of detailed observations
from several SGRs \citep{Israel2005,SW2005,WS2006} revealed that the
decaying part of the spectrum exhibits a number of oscillations with
frequencies in the range of a few tenths of Hz to a few hundred Hz.
There are three events detected up to now which can be associated
with crust oscillations of magnetars. The first event was detected
already in 1979 from the source SGR 0526-66
\citep{Mazets1979,Barat1983}, the second in 1998 from SGR 1900+14
\citep{Hurley1999}, while the third and most energetic one was
observed in December 2004 from the source SGR 1802-20
\citep{Terasawa2005, Palmer2005}. Analysis of the tail oscillations
of SGR 1806-20 revealed the presence of oscillations at
 approximately 18, 26, 29, 92.5,
150, 626.5, and 1837 Hz \citep{Israel2005, WS2006,SW2006},
for SGR 1900+14 the
detected frequencies were 28, 54, 84, and 155 Hz \citep{SW2005},
while for the case of SGR 0526-66 there was only one frequency
identified at 43.5 Hz \citep{Barat1983}.


Magnetar starquakes may be driven by the evolving intense magnetic
field, which accumulates stress and eventually leads to crust
fracturing.  The excited seismic oscillations are known to be of two
types: {\em shear modes}, which are polar-type oscillations and {\em
  torsional modes}, which are axial-type oscillations. The latter are
thought to be more easily excited during a fracturing of the crust,
since they only involve oscillations of the velocity, with the shear
modulus of the crust acting as the necessary restoring force. Slowly
rotating stars still maintain a nearly spherical shape and the velocity
field of torsional oscillations is then divergence-free, with no
radial component. Torsional modes ({\em t}-modes) are labeled as
$_\ell t_n$, where $\ell$ is the angular index, while the index $n$
corresponds to the number of radial nodes in the eigenfunctions of the
overtones for a specific $\ell$. Shear and torsional modes have been
studied mainly in Newtonian theory, see e.g.
\citet{HC1980,McDermott1988,Carroll1986,
  Strohmayer1991a,Lee2006}, while there are only a few studies of torsional modes
in general relativity \citep{Schumaker1983, Leins1994,
Messios2001,SA2006}.

In the Newtonian, non-magnetized limit \citet{HC1980}
found that the period
$_\ell P_0 $ of the fundamental torsional modes depends mostly on
the radius of the star, $R$, the speed of shear waves, $u_s$, and
the angular index $\ell$, via
\begin{equation}
{}_\ell P_0 \approx 2\pi[\ell(\ell+1)]^{-1/2} R/u_s, 
\end{equation}
where  $u_s=(\mu/\rho)^{1/2}$, with $\mu$ and $\rho$ being the
shear modulus and the density, respectively. Furthermore, it was found
that the period of overtones of index $n$ is essentially independent
of the angular index $l$ (provided $l$ does not greatly exceed $n$)
and is basically determined by the crust thickness $\Delta r$:
\begin{equation}
{}_\ell P_n \approx 2n^{-1}\Delta r/u_s.
\end{equation}
From the above two relations it follows that
the relative crust thickness, $\Delta r/R$, is given by
\begin{equation}
\frac{\Delta r}{R} \approx \frac{\pi n}{\sqrt{\ell (\ell+1)}}
 \frac{_\ell f_0}{_\ell f_n},
 \label{eq:mode_spacing}
\end{equation}
which is independent of the details of the equation of state in the
crust and basically only depends on the ratio of the frequency of the
overtones to the fundamental frequency, see also \citet{SA2006}. It
thus becomes obvious that the successful identification of both the
fundamental torsional frequencies and their overtones could allow for
the determination of the crust thickness, which in turn could lead to
information on the high-density part of the equation of state. In such
a determination, of course, general relativistic and magnetic field
effects on the frequencies of torsional modes should properly be taken
into account, which is one of the aims of the present paper.

Relativistic effects have been found to significantly increase the
fundamental $\ell=2$ torsional mode period by roughly 30\% for
a typical $M=1.4M_\odot$, $R=10{\rm km}$ model. \citet{Duncan1998}
derived the following empirical formula for the period of the
fundamental $\ell=2$ torsional mode of a relativistic star
\begin{equation}
_\ell P_0 ({\rm ms}) \approx 33.6 R_{\rm 10}\frac{0.87+0.13M_{\rm 1.4}
R^{-2}_{\rm 10}}{(1.71-0.71M_{\rm 1.4} R^{-1}_{\rm 10})^{1/2}},
\label{eq:Duncan}
\end{equation}
where $R_{\rm 10}=R/(10 {\rm km})$ and $M_{\rm 10}=M/(1.4M_\odot)$.
 In this relation, two effects have been taken into account, 
the (relativistic) gravitational redshift and the (Newtonian) acceleration 
  of gravity $g=GM/R^2$ at the surface. 
  More details can be found in \cite{Duncan1998}, see also \cite{Messios2001}.
In addition, the formula (\ref{eq:Duncan}) 
agrees well with the fully relativistic empirical relation found
by \citet{Messios2001}
\begin{equation}
{}_{\ell} P_0 ({\rm ms}) \approx 34 R_{\rm 10}.
\end{equation}
The latter is an extension of the Newtonian Hansen \& Cioffi formula, based
on a particular $1.4 M_\odot$ model. On the other hand, the effect of a
strong magnetic field, $B$, on torsional modes was
considered by \citet{Duncan1998} and \citet{Messios2001}. Assuming
that the shear modulus $\mu$ is augmented by the magnetic field tension,
$B^2/4\pi$, one can derive the following estimate for the period
of a torsional mode
\begin{equation}
P \approx P^{(0)}\left[1+(B/B_\mu)\right]^{-1/2}, \label{Pb}
\end{equation}
where $P^{(0)}$ is the period in the limit of vanishing magnetic field
and $B_\mu=\sqrt{4\pi\mu}$. However, \citet{Messios2001} found that
if the structure of the magnetic field is close to monopolar in the
crust, the period of the fundamental $\ell=2$ torsional mode does
not follow the above relation. This underlines the importance of
properly taking into account the structure of the magnetic field, so
that Equation (\ref{Pb}) must be generalized to particular
magnetic field multipoles.

An essential ingredient in computing torsional mode frequencies is the
assumption one makes about the shear modulus $\mu$. Here, we adopt
the zero-temperature limit of the approximate formula derived
by \citet{Strohmayer1991b}
\begin{equation}
\mu= 0.1194 \frac{n_i (Ze)^2}{a},
\label{eq:mu}
\end{equation}
where $n_i$ is the ion number density, $a^3=3/4\pi n_i$ is the average
ion spacing, and $+Ze$ is the ion charge.
Notice that this approximate formula for the shear modulus is derived
  on the assumption of a bcc crystalline state, which is averaged over all directions.
For the detailed composition
of the crust, we consider a modern EOS by \citet{DH2001} (hereafter DH)
and, for comparison, an older EOS by \citet{NV1973} (hereafter NV). The
two EOSs for the crust differ both in composition as well as in the
value of the density at the base of the crust. As the crust thickness
depends on the compactness of the star, we explore
the allowed parameter space for the unknown high-density EOS by
selecting several tabulated high-density EOSs that yield neutron star
models with significantly different macroscopic properties.

In the present work, we derive the perturbation equations for
torsional oscillations of magnetized stars in general relativity
following the approach by \citet{Messios2001} and neglecting spacetime
perturbations (Cowling approximation). We specialize the magnetic
field to a dipole structure, which allows us to reduce the system of
equations to only one dimension. A large number of torsional modes
(fundamental and overtones) is obtained for selected magnetar models
differing in the high-density EOS, the crust EoS and mass. Our study
is valid for moderate magnetic field strengths, up to roughly
$10^{16}$ G, assuming that the magnetic field is confined to the crust
and neglecting the possible presence of a thin fluid ocean.  The
reason for this is twofold: on one hand, for significantly stronger
magnetic fields the distortion of the equilibrium shape of the star as
well the magnetic coupling to higher harmonics should also be taken
into account; on the other hand, the Alfv\'en velocity $u_A\equiv
B/(4\pi\rho)^{1/2}$ $\approx 3\times 10^7$ cm/s $B_{15} \,
\rho_{14}^{-1/2}$ becomes comparable to the speed of shear waves in
the crust when the magnetic field reaches values of
10$^{15}$-10$^{16}$G, which implies that global magnetosonic waves
will be strongly coupled to shear waves and the torsional oscillations
may no longer be confined to the crust. The effect of including global
magnetosonic waves even at smaller magnetic field strengths should
also be studied in more detail, see \citep{GSA2006,SKSV06}.

The article is structured as follows: in the next two sections we
describe the general-relativistic ideal MHD equations, the
background stellar configuration and the way that we have derived the
dipole geometry of the magnetic field. In the fourth section we derive
the perturbation equations for torsional oscillations of magnetized
stars in the Cowling approximation, while in the fifth section we
present the obtained frequencies for selected magnetar models along
with fitting formulae for the influence of the magnetic field in the
oscillation spectrum. The article closes with a summary and discussion.

Unless otherwise noted, we adopt
units of $c=G=1$, where $c$ and $G$ denote the speed of light and
the gravitational constant, respectively, while the metric signature is
$(-,+,+,+)$.

\section{General-Relativistic Ideal MHD}
\label{sec:II}

The energy momentum tensor $T^{\mu\nu}$ for a magnetized
relativistic star in equilibrium, is the sum of the stress-energy
tensors of a perfect fluid $T^{\mu\nu(pf)}$ and the magnetic field
$T^{\mu\nu(M)}$
\begin{equation}
 T^{\mu\nu} = T^{\mu\nu(pf)} + T^{\mu\nu(M)},
\label{t_munu} \,
\end{equation}
where
\begin{eqnarray}
 T^{\mu\nu(pf)} &=& (\epsilon + p)u^{\mu}u^{\nu} + pg^{\mu\nu}, \\
 T^{\mu\nu(M)}  &=& H^2 u^{\mu} u^{\nu} + \frac{1}{2} H^2 g^{\mu\nu}
- H^{\mu} H^{\nu}. \,
\end{eqnarray}
Above, $\epsilon$, $p$, and $H^{\mu}$ are the energy density, the
pressure, and a normalized magnetic field, which we define by absorbing
a factor of $\sqrt{4\pi}$, i.e., $H^{\mu}\equiv B^{\mu}/\sqrt{4\pi}$, and
$H^2 = H_{\mu}H^{\mu}$. Here and throughout the paper we will assume the
{\it ideal MHD} approximation. In addition, we assume the shear modulus
to be isotropic, so that there is no contribution of the shear stress
in equilibrium.

Using the stress-energy tensor (\ref{t_munu}), one obtains the
equations of motion of the fluid by projecting the conservation of
the energy-momentum tensor on to the hypersurface normal to $u^\mu$,
i.e.,
\begin{equation}
h^{\mu}_{\ \alpha}T^{\alpha\nu}_{\ \ ;\nu}=0 \, ,
\end{equation}
where $h^{\mu}_{\ \nu}= g^{\mu}_{\ \nu}+u^{\mu}u_{\nu}$ is the projection
tensor. The equations of motion are
\begin{equation}
 (\epsilon + p + H^2) u^{\mu}_{\ ;\nu}u^{\nu} = -h^{\mu\nu}
\left(p+\frac{1}{2}H^2\right)_{;\nu}
     + h^{\mu}_{\ \alpha}\left(H^{\alpha}H^{\nu}\right)_{;\nu}
     .  \label{conservation}
\end{equation}

Due to the ideal MHD approximation the electric field 4-vector
vanishes, $E_{\mu}=F_{\mu\nu}u^{\nu}=0$ so that the electric
field is zero for a comoving observer.  In this
approximation, the Maxwell's equations $F_{[\mu\nu;\gamma]}=0$, where
$F_{\mu\nu}$ is the Faraday tensor, can be written in the simple form:
\begin{equation}
 (u^{\mu} H^{\nu} - u^{\nu} H^{\mu})_{;\mu}=0. \label{Maxwell0-1}
\end{equation}
In deriving the above equation the following definitions are used
\begin{eqnarray}
 F_{\mu\nu} &=& u_{\mu}E_{\nu} - u_{\nu}E_{\mu} -
\epsilon_{\mu\nu\alpha\beta}u^{\alpha}B^{\beta}, \\
 B_{\mu} &=& \frac{1}{2} \epsilon_{\mu\nu\alpha\beta} u^{\nu}
F^{\alpha\beta}, \label{magnetic-field}
\end{eqnarray}
From Equation (\ref{Maxwell0-1}) one derives the
magnetic induction equation
\begin{eqnarray}
 H^{\mu}_{\ ;\nu}u^{\nu} &=& -u^{\alpha}_{\ ;\alpha}H^{\mu}
+ u^{\mu}_{\ ;\nu}H^{\nu}
     + H^{\alpha}u_{\alpha;\beta}u^{\beta}u^{\mu}, \label{Maxwell1} \\
                         &=& \left(\sigma^{\mu}_{\ \nu} + \omega^{\mu}_{\ \nu}
     - \frac{2}{3}\delta^{\mu}_{\ \nu}\Theta\right)H^{\nu}
     + H^{\alpha}u_{\alpha;\beta}u^{\beta}u^{\mu}, \label{Maxwell2}
\end{eqnarray}
where $\sigma_{\mu\nu}$ is the rate of shear tensor, $\omega_{\mu\nu}$ is the
twist tensor and $\Theta$ is the expansion, defined as
\begin{eqnarray}
 \Theta &\equiv& u^{\mu}_{\ ;\mu}, \\
 \sigma_{\mu\nu} &\equiv& \frac{1}{2}\left(u_{\mu;\alpha}h^{\alpha}_{\ \nu}
     + u_{\nu;\alpha}h^{\alpha}_{\ \mu}\right)
- \frac{1}{3}\Theta h_{\mu\nu}, \\
 \omega_{\mu\nu} &\equiv& \frac{1}{2}\left(u_{\mu;\alpha}h^{\alpha}_{\ \nu}
     - u_{\nu;\alpha}h^{\alpha}_{\ \mu}\right).
\end{eqnarray}
In a following section, the above equations will be linearized, in
order to obtain the perturbed equations of motion describing
small-amplitude oscillations. But first, we describe the construction
of the background equilibrium models.

\section{Equilibrium Configuration}

A strongly magnetized relativistic star has a non-spherical shape,
due to the tension of the magnetic field, which is expressed via the
off-diagonal components in the stress-energy tensor $T^{\mu\nu}$.
However, the deformations from spherical symmetry induced by the
magnetic field are small for neutron stars with magnetic field
usually assumed for magnetars, because the energy of the magnetic
field is considerably smaller then the gravitational energy, i.e.,
\begin{equation}
 \frac{{\rm magnetic\ energy}}{{\rm gravitational\ energy}}
    \sim \frac{B^2 R^3}{G M^2 / R}
    \sim 10^{-4} \left(\frac{B}{10^{16} [{\rm G}]}\right)^2.
\end{equation}
For this reason, we neglect the deformations due to magnetic fields
in the construction of equilibrium models. Since magnetars are also
extremely slowly rotating (the intense magnetic field spins down the
star on a short timescale) we can also neglect any rotational deformations,
so that the equilibrium models can be considered as spherically
symmetric solutions of the well-known TOV equations described by
a metric of the form
\begin{equation}
 ds^2 = -e^{2\Phi}dt^2 + e^{2\Lambda}dr^2 + r^2(d \theta^2 + \sin^2\theta
d \phi^2), \label{metric}
\end{equation}
where $\Phi$ and $\Lambda$ are the function of the Schwarzschild
radial coordinate $r$.  The  four-velocity of the equilibrium model
is thus
\begin{equation}
u^{\mu} = (e^{-\Phi},0,0,0).
\end{equation}

We supplement the equilibrium model by a dipole magnetic field (but
see \citet{BS2006} for other assumptions), following the approach of
\citet{Konno1999} i.e. we consider an axisymmetric,
poloidal magnetic field, which is created by a 4-current
$J_{\mu}=(0,0,0,J_{\phi})$.  In ideal MHD the electromagnetic
4-potential $A_{\mu}$ is very simple and has only one component in
spherical polar coordinates
\begin{equation}
 A_{\mu} = (0,0,0,A_{\phi}),
\end{equation}
with
\begin{equation}
F_{\mu\nu} =
A_{\nu,\mu} - A_{\mu,\nu},
\end{equation}
 \citep{Bocquet1995,Konno1999}. For the metric chosen above,
Maxwell's equations $F^{\mu\nu}_{\ \
  ;\nu}=4\pi J^{\mu}$ lead to the following elliptic equation for
$A_{\phi}$
\begin{equation}
 e^{-2\Lambda} \frac{\partial^2 A_{\phi}}{\partial r^2}
+ \frac{1}{r^2}\frac{\partial^2 A_{\phi}}{\partial \theta^2}
     + \left(\Phi' - \Lambda' \right)e^{-2\Lambda}
\frac{\partial A_{\phi}}{\partial r}
     - \frac{1}{r^2} \frac{\cos\theta}{\sin\theta}
\frac{\partial A_{\phi}}{\partial \theta}
     = -4\pi J_{\phi}\, . \label{Maxwell-J}
\end{equation}
The 4-vectors $A_{\mu}$ and $J_{\mu}$ can be expanded in
vector spherical harmonics, so that
\begin{eqnarray}
 A_{\phi}(r,\theta) &=& a_{\ell_M}(r) \sin\theta
\partial_{\theta}P_{\ell_M}(\cos\theta), \\
 J_{\phi}(r,\theta) &=& j_{\ell_M}(r) \sin\theta
\partial_{\theta}P_{\ell_M}(\cos\theta),
\end{eqnarray}
where $a_{\ell_M}(r)$ and $j_{\ell_M}(r)$ are functions of the radial
coordinate only. This leads to a one-dimensional form of Equation
(\ref{Maxwell-J})
\begin{equation}
 e^{-2\Lambda} \frac{d^2 a_{\ell_M}}{dr^2} + \left(\Phi' - \Lambda' \right)
e^{-2\Lambda} \frac{d a_{\ell_M}}{dr}
     - \frac{\ell_M(\ell_M+1)}{r^2} a_{\ell_M} = -4\pi j_{\ell_M}.
\label{Maxwell-J1}
\end{equation}

\begin{figure}
\includegraphics[width=65mm]{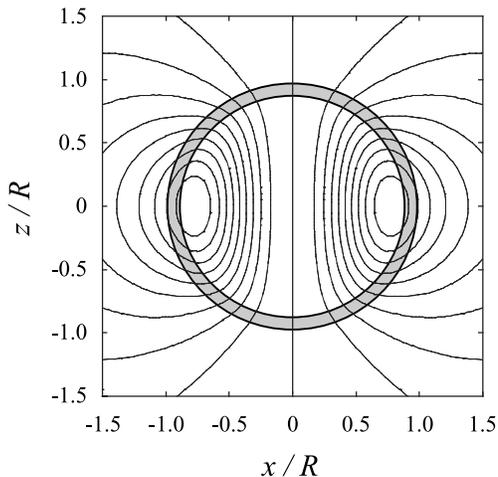}
 \caption{Magnetic field lines of a typical equilibrium model. At the
   surface, an analytic solution for an exterior dipole magnetic field
   is smoothly joined to a numerical interior solution. The grey
   area shows the crust region, to which the magnetic field is 
   assumed to be effectively confined when computing the torsional
   oscillations.}
  \label{Fig:dipole_field}
\end{figure}

For a dipole magnetic field, i.e., $\ell_M=1$, an
analytic exterior solution for $a_1$ exists, setting
$j_1^{\rm (ex)}=0$ and $e^{2\Phi} = e^{-2\Lambda} = 1-2M/r$
\citep{Wasserman1983}
\begin{equation}
 a_1^{\rm (ex)} = -\frac{3\mu_b}{8M^3}r^2 \left[\ln\left(1-\frac{2M}{r}\right)
 + \frac{2M}{r} + \frac{2M^2}{r^2}\right],
\end{equation}
where $\mu_b$ is the magnetic dipole moment for an observer at infinity.
In the interior the potential $a_1^{\rm (in)}$ is found by solving
numerically Equation (\ref{Maxwell-J1}).  The current distribution
$j_1^{\rm (in)}$, which is the source term of Equation
(\ref{Maxwell-J1}) must satisfy the integrability condition
that the solution is stationary
\citep{Bonazzola1993}.  Here we adopt the same, simplified form for
the current distribution as the one used in \citet{Konno1999}, i.e.,
\begin{equation}
 j_1^{(\rm in)} = f_0 r^2 (\epsilon + p),
\end{equation}
where $f_0$ is an arbitrary constant.  Moreover, regularity at the
center of the star implies that $a_1^{\rm (in)}$ has the form
\begin{equation}
 a_1^{\rm (in)} = \alpha_c r^2 + {\cal O}(r^4),
\end{equation}
where $\alpha_c$ is a constant, which is determined by the junction
condition with $a_1^{\rm (ex)}$ at stellar surface.  Finally, the
components of the vector $H_\mu$ describing the normalized magnetic
field are given by the following set of equations
\begin{eqnarray}
 H_r &=& -\frac{e^{\Lambda}}{\sqrt{4\pi}r^2\sin\theta}\partial_{\theta}A_{\phi}
      = \frac{e^{\Lambda}\cos\theta}{\sqrt{\pi}r^2}a_1\, , \label{eq:H_r} \\
 H_{\theta} &=& \frac{e^{-\Lambda}}{\sqrt{4\pi}\sin\theta}\partial_r A_{\phi}
             = -\frac{e^{-\Lambda}\sin\theta}{\sqrt{4\pi}} {a_1}_{,r}.
\label{eq:H_theta}
\end{eqnarray}
Figure  \ref{Fig:dipole_field} shows the magnetic field lines of the
complete solution, for a typical equilibrimum configuration. 
 We emphasize that, even though the magnetic field configuration has
been computed throughout the star, the boundary conditions we will
assume for the torsional oscillations will neglect the
presence of the magnetic field outside the crust region, so that, 
effectively, that magnetic field will be assumed to be confined
to the crust only.

\section{Perturbation Equations}
\label{sec:III}

\subsection{Covariant form}
The perturbation equations are derived by linearizing the equations of
motion (\ref{conservation}) and the magnetic induction equation
(\ref{Maxwell1}). Since we only consider torsional oscillations, which
are of axial type and do not induce density variations in spherical
stars, there is no significant variation in the radiative part of
the metric describing the gravitational field.  In addition, since
torsional modes are essentially material velocity oscillations, the
imaginary part of their frequency due to the emission of
current-multipole gravitational radiation is much smaller than the
real part.  For these reasons the frequency of torsional oscillations
is determined with satisfactory accuracy even when neglecting entirely
the metric perturbations by setting $\delta g_{\mu\nu}=0$ (i.e.
adopting the relativistic Cowling approximation), see
\citet{Leins1994} and we adopt this approximation here.

The linearized form of the equation of motion (\ref{conservation}) is
\begin{eqnarray}
 (\epsilon + p + H^2) \delta u^{\mu}_{\ ;\nu}u^{\nu}
    &=& -(\delta \epsilon + \delta p + 2H_{\alpha}\delta
H^{\alpha})u^{\mu}_{\ ;\nu}u^{\nu}
    - (\epsilon + p + H^2)u^{\mu}_{\ ;\nu}\delta u^{\nu} \nonumber \\
   &+& (u^{\mu}\delta u_{\alpha} + \delta u^{\mu}u_{\alpha})
      \left[H^{\alpha}H^{\nu} - g^{\alpha\nu}\left(p+\frac{1}{2}H^2\right)
\right]_{;\nu} \nonumber \\
   &+& h^{\mu}_{\ \alpha}\left[H^{\alpha} \delta H^{\nu} +
\delta H^{\alpha}H^{\nu}
    - g^{\alpha\nu}\left(\delta p + H_{\beta}\delta
H^{\beta}\right)\right]_{;\nu}
    - h^{\mu}_{\ \alpha}\delta T^{\alpha\nu(s)}_{\ \ ;\nu}.
\label{perturbation1}
\end{eqnarray}
where $\delta T^{\mu\nu(s)}$ is the linearized shear stress
tensor. The latter is assumed to be related  to the
linearized shear tensor $\delta S_{\mu\nu}$, through
\begin{equation}
\delta T^{(s)}_{\mu\nu} = -2\mu \delta S_{\mu\nu},
\end{equation}
see \citet{Schumaker1983}. In addition, linearization of the
 magnetic induction equation
(\ref{Maxwell1}) yields
\begin{eqnarray}
 \delta H^{\mu}_{\ ;\nu}u^{\nu}
     &=& -H^{\mu}_{\ ;\nu}\delta u^{\nu}
     + h^{\mu\alpha}\delta u_{\alpha;\beta}H^{\beta}
     + h^{\mu\alpha}u_{\alpha;\beta}\delta H^{\beta}
     + u^{\mu}\delta u^{\alpha}H^{\beta}u_{\alpha;\beta}
     - \delta \Theta H^{\mu} - \Theta \delta H^{\mu} \nonumber \\
    &+& u^{\mu} H^{\alpha}\left(\delta u_{\alpha;\beta}u^{\beta}
+ u_{\alpha;\beta}\delta u^{\beta}\right)
     + \left(H^{\alpha}\delta u^{\mu}
+ \delta H^{\alpha}u^{\mu}\right)u_{\alpha;\beta}u^{\beta} \, ,
     \label{perturbation2}
\end{eqnarray}
where we used $u^{\alpha}\delta u_{\alpha;\beta} =-\delta
u^{\alpha}u_{\alpha;\beta}$. Notice that the above relation differs in
some terms from Equation (12) in \citet{Messios2001}, due to a sign
error in the latter reference\footnote{A sign error adopted initially
  from another publication propagated throughout the magnetic terms in
  the equations and final numerical results presented in that paper.
  We have confirmed that all equations and results in the non-magnetized
  limit were correct, while also the qualitative behaviour of the
  fundamental torsional mode as a function of the magnetic field
  strength was correct (see Appendix \ref{sec:IV}).}. 
  In Appendix \ref{sec:IV} we also give the
corrected equations and numerical results for the special magnetic
field configuration considered in \citet{Messios2001}.

\subsection{Axial Perturbations}

The usual method for studying matter perturbations of spherically
symmetric backgrounds is to decompose the perturbed quantities into
scalar and vector spherical harmonics of definite indices $\ell$ and
$m$. In this way, the different oscillation modes characterized by
$\ell$ and $m$ decouple and one can study them independently. Here
we restrict attention to axial-type perturbations, for which the
perturbed matter quantities can be written as
\begin{eqnarray}
 \delta \epsilon &=& \delta p = 0, \\
 \delta u^{t} &=& \delta u^{r} = \delta u^{\theta} = 0, \\
 \delta u^{\phi} &=& e^{-\Phi} \partial_t {\cal Y}(t,r) b(\theta),
\end{eqnarray}
where
\begin{equation}
 b(\theta) \equiv \frac{1}{\sin\theta}\partial_{\theta}
P_{\ell}(\cos\theta)\, ,
\end{equation}
where $\partial_\theta$ denotes the partial derivative with respect to $\theta$.
Above, ${\cal Y}(t,r)$ describes the radial dependence of the angular
displacement of the stellar material, $P_{\ell}(\cos\theta)$ is the
Legendre polynomial of order $\ell$ and we have set $m=0$, due to
the degeneracy in $m$ for spherically symmetric backgrounds. The above
assumptions lead to the following form of the $\phi$-component of the
linearized equation of motion (\ref{perturbation1})
\begin{eqnarray}
 (\epsilon + p + H^2)e^{-\Phi} \delta u^{\phi}_{,t} &=&
       \delta H^r \left[H^{\phi}_{\ ,r} + H^{\phi}\left(\Phi_{,r}
+ \Lambda_{,r} + \frac{4}{r}\right)
     - \frac{e^{2\Lambda}H^r_{\ ,\phi}}{r^2\sin^2\theta}\right]
     + \delta H^{\theta} \left[H^{\phi}_{\ ,\theta}
+ 3\cot\theta H^{\phi}
     - \frac{H^{\theta}_{\ ,\phi}}{\sin^2\theta}\right] \nonumber \\
    &+& \delta H^{\phi} \left[2\cot\theta H^{\theta}
+ \left(\Phi_{,r} + \frac{2}{r}\right)H^r\right]
     + H^{\phi}\left(\delta H^t_{\ ,t} + \delta H^r_{\ ,r}
+ \delta H^{\theta}_{\ ,\theta} + \delta H^{\phi}_{\ ,\phi}\right)
     + H^r \delta H^{\phi}_{\ ,r} +  H^{\theta}\delta H^{\phi}_{\ ,\theta}
\nonumber \\
    &-&
 \frac{H^r \delta H_{r,\phi}}{r^2\sin^2\theta}
     - \frac{H^{\theta}\delta H_{\theta,\phi}}{r^2\sin^2\theta}
     - \delta T^{r\phi(s)}_{\ \ ,r} - \delta T^{\theta\phi(s)}_{\ \ ,\theta}
     - \left(\frac{4}{r} + \Phi_{,r} + \Lambda_{,r}\right)\delta T^{r\phi(s)}
     - 3\cot\theta \delta T^{\theta\phi(s)},
\label{perturbation-equation1}
\end{eqnarray}
where we used the relation  $H^{\alpha}_{\ ;\alpha} =
u^{\alpha}_{\ ;\beta}u^{\beta}H_{\alpha}$, which is derived from the
Maxwell's equations.

For the derivation of the linearized shear tensor $\delta S_{\mu\nu}$,
we adopt the relationship $\delta\sigma_{\mu\nu} = {\cal L}_{u}\delta
S_{\mu\nu}$ \citep{ CQ1972,Schumaker1983}, which becomes
\begin{equation}
\delta \sigma_{\mu\nu} = e^{-\Phi} \delta S_{\mu\nu,t}
\end{equation}
In the Cowling approximation the only nonvanishing components of
$\delta\sigma_{\mu\nu}$ are written as
\begin{eqnarray}
 \delta \sigma_{r\phi} = \frac{1}{2} r^2 e^{-\Phi} {\cal Y}_{,tr}
b\sin^2\theta \hspace{1cm}
 \mbox{and} \hspace{1cm}
 \delta \sigma_{\theta\phi} = \frac{1}{2} r^2 e^{-\Phi} {\cal Y}_{,t}
b_{,\theta} \sin^2\theta ,
\end{eqnarray}
It follows that the only non-vanishing
components of the linearized shear stress tensor are
\begin{eqnarray}
 \delta T^{(s)}_{r\phi} &=& -\mu r^2 {\cal Y}_{,r} b \sin^2\theta, \\
 \delta T^{(s)}_{\theta\phi} &=& -\mu r^2 {\cal Y} b_{,\theta} \sin^2\theta.
\end{eqnarray}
The linearized induction equation (\ref{perturbation2}) yields that
following relations for the first time-derivative of the components of
 $\delta H^\mu$
\begin{eqnarray}
 \delta H^t_{\ ,t} &=& e^{-\Phi}H^{\phi}\delta u_{\phi,t}, \\
 \delta H^r_{\ ,t} &=& -e^{\Phi}H^r_{\ ,\phi} \delta u^{\phi}, \\
 \delta H^{\theta}_{\ ,t} &=& -e^{\Phi} H^{\theta}_{\ ,\phi}
\delta u^{\phi}, \\
 \delta H^{\phi}_{\ ,t} &=& e^{\Phi}\left[\left(\Phi_{,r}H^r
- H^{\phi}_{\ ,\phi}\right)\delta u^{\phi}
     + H^r \delta u^{\phi}_{\ ,r} + H^{\theta}\delta u^{\phi}_{\ ,\theta}
\right].
\end{eqnarray}
Substituting the previous relations into the linearized equations
of motion (\ref{perturbation-equation1}) and  assuming that all
perturbed variables have a harmonic time dependence, such
that ${\cal Y}(t,r)=e^{i \omega t}{\cal Y}(r)$, one obtains
\begin{eqnarray}
 -\left[\epsilon + p + H^2 - \left(H^{\phi}r\sin\theta\right)^2 \right]
 \omega^2 e^{-2\Phi} {\cal Y}
    &=&   e^{-2\Lambda}\mu {\cal Y}''
     + \left[\left(\frac{4}{r} + \Phi' - \Lambda' \right)\mu +
\mu'\right]e^{-2\Lambda}{\cal Y}'
     - \frac{(\ell+2)(\ell-1)}{r^2}\mu {\cal Y} \nonumber \\
    &-& \Bigg\{H^r_{\ ,\phi} H^{\phi}_{\ ,r} + H^r_{\ ,\phi} H^{\phi}
\left(\Phi' + \Lambda' + \frac{4}{r}\right)
     + H^{\theta}_{\ ,\phi} H^{\phi}_{\ ,\theta}
     + 3\cot\theta H^{\theta}_{\ ,\phi} H^{\phi}
\nonumber \\
     &-& \frac{1}{r^2\sin^2 \theta}e^{2\Lambda}\left[(H^r_{\ ,\phi})^2
+ H^rH^r_{\ ,\phi\phi}\right]
     - \frac{1}{\sin^2\theta}\left[(H^{\theta}_{\ ,\phi})^2
+ H^{\theta}H^{\theta}_{\ ,\phi\phi}\right] \nonumber \\
     &+& H^{\phi}_{\ ,\phi} \left[2\cot\theta H^{\theta}
     + \left(\Phi' + \frac{2}{r}\right)H^r\right]
     + H^{\phi}\left(H^r_{\ ,r\phi} + H^{\theta}_{\ ,\theta\phi}
+ H^{\phi}_{\ ,\phi\phi}\right)
     + H^rH^{\phi}_{\ ,r\phi} \nonumber \\
 &+& H^{\theta}H^{\phi}_{\ ,\theta\phi}\Bigg\}
{\cal Y}
    + \left[2\cot\theta H^rH^{\theta} + \left(\Phi'
+ \frac{2}{r}\right)(H^r)^2
     + H^r \left(H^r_{\ ,r} - H^{\phi}_{\ ,\phi}\right)
     + H^{\theta}H^r_{\ ,\theta}\right] {\cal Y}' \nonumber \\
   &+& (H^r)^2  {\cal Y}''
    +\Bigg\{\left[\left(\Phi' + \frac{2}{r}\right)H^rH^{\theta}
+ H^rH^{\theta}_{\ ,r}
     + 2\cot\theta (H^{\theta})^2
     + H^{\theta} \left(H^{\theta}_{\ ,\theta}
- H^{\phi}_{\ ,\phi}\right)\right]{\cal Y} \nonumber \\
     &+& 2H^rH^{\theta}{\cal Y}'\Bigg\}\frac{b_{,\theta}}{b}
    + \left(H^{\theta}\right)^2 {\cal Y} \frac{b_{,\theta\theta}}{b},
\label{eigen}
\end{eqnarray}
where a prime $(')$ denotes the derivative with respect to $r$. The
above eigenvalue equation for the mode-frequency $\omega$ is still
written for a general equilibrium magnetic field $H^\mu(r,\theta,\phi)$.

\subsection{Dipole magnetic field}

Restricting attention to an axisymmetric poloidal magnetic field,
the eigenvalue equation (\ref{eigen}) becomes
\begin{eqnarray}
 -\big[\epsilon + p + H^r H_r + H^{\theta}H_{\theta}\big]\omega^2 e^{-2\Phi}
 {\cal Y}
    &=& \ e^{-2\Lambda}\mu {\cal Y}''
     + \left[\left(\frac{4}{r} + \Phi' - \Lambda' \right)\mu
+ \mu'\right]e^{-2\Lambda}{\cal Y}'
     - (\ell+2)(\ell-1) \left[\frac{\mu}{r^2} + (H^{\theta})^2\right] {\cal Y}
 \nonumber \\
    &+& \left[2\cot\theta H^rH^{\theta} + \left(\Phi'
+ \frac{2}{r}\right)(H^r)^2
     + H^rH^r_{\ ,r} + H^{\theta}H^r_{\ ,\theta}\right] {\cal Y}' + (H^r)^2
 {\cal Y}'' \nonumber \\
    &+&\left\{\left[\left(\Phi' + \frac{2}{r}\right)H^rH^{\theta}
+ H^rH^{\theta}_{\ ,r}
     - \cot\theta (H^{\theta})^2
+ H^{\theta}H^{\theta}_{\ ,\theta}\right]{\cal Y}
     + 2H^rH^{\theta}{\cal Y}'\right\}\frac{b_{,\theta}}{b},
\label{master-equation}
\end{eqnarray}
where we used the following equation satisfied by the angular function
$b(\theta)$
\begin{equation}
 b_{,\theta\theta} + 3\cot \theta b{,_\theta} + (\ell+2)(\ell-1)b = 0 \, .
\end{equation}
The above eigenvalue equation can be simplified further by assuming
a dipole magnetic field of the form (\ref{eq:H_r}), (\ref{eq:H_theta}).
We achieve this by first writing Equation (\ref{master-equation}) in
the form
\begin{eqnarray}
 [{\cal A}_\ell(r) + {\cal B}_\ell(r)] \partial_{\theta}Y_{\ell 0}
     + [{\cal C}_\ell(r) - {\cal B}_\ell(r)] \sin^2 \theta
\partial_{\theta}Y_{\ell 0}
     + {\cal D}_\ell(r) \sin \theta \cos \theta Y_{\ell 0} =0,
     \label{master-equation1}
\end{eqnarray}
where $Y_{\ell 0}$ is the spherical harmonic function with $m=0$,
which is related to $P_\ell(\cos\theta)$ via the normalization
$Y_{\ell0} = \sqrt{(2\ell+1)/(4\pi)} P_\ell$.
The four
radial functions ${\cal A}_\ell(r)$, ${\cal B}_\ell(r)$, ${\cal
  C}_\ell(r)$, and ${\cal D}_\ell(r)$ are
\begin{eqnarray}
 {\cal A}_\ell(r) &=& \mu {\cal Y}'' + \left[\left(\frac{4}{r} + \Phi'
- \Lambda' \right)\mu + \mu' \right] {\cal Y}'
     + \left[\left(\epsilon + p \right) \omega^2 e^{-2\Phi}
     - \frac{(\ell+2)(\ell-1)}{r^2}\mu\right]e^{2\Lambda} {\cal Y}, \\
 {\cal B}_\ell(r) &=&  \frac{a_1}{\pi r^4} \left\{a_1 {\cal Y}''
     + \left[\left(\Phi' - \Lambda' \right)a_1  + 2{a_1}' \right] {\cal Y}'
     + \left[\omega^2 e^{-2\Phi + 2\Lambda} a_1 + \left(\Phi'
- \Lambda' \right){a_1}'
     + {a_1}'' \right] {\cal Y}\right\}, \\
 {\cal C}_\ell(r) &=& \frac{{a_1}'}{4\pi r^4}\left\{2a_1 {\cal Y}'
     + \left[\omega^2 e^{-2\Phi}r^2 - (\ell+2)(\ell-1)\right]
{a_1}' {\cal Y} \right\}, \\
 {\cal D}_\ell(r) &=& \frac{\ell(\ell+1)}{2\pi r^4}{a_1}\left\{2{a_1}'
     {\cal Y}'
     + \left[\left(\Phi' - \Lambda' \right){a_1}' + {a_1}''\right]
{\cal Y} \right\},
\end{eqnarray}
in which the magnetic field now appears through the radial function
$a_1$ of Equations (\ref{eq:H_r}) and (\ref{eq:H_theta}).

Multiplying Equation (\ref{master-equation1}) with
$\partial_{\theta}Y^*_{\ell0}$ and integrating over the sphere, the
eigenvalue equations takes the 1-dimensional form
\begin{eqnarray}
 \ell(\ell+1)\left({\cal A}_\ell + {\cal B}_\ell\right) +{\cal L}^{\pm 2}_1
\left({\cal C}_\ell - {\cal B}_\ell\right)
     + {\cal L}^{\pm 2}_2 {\cal D}_\ell = 0\, . \label{system}
\end{eqnarray}
In deriving this equation we used the following properties of
spherical harmonic functions:
\begin{eqnarray}
\ell(\ell+1) {\cal A}_\ell &:= & \sum_{\ell'}{\cal A}_{\ell'}
\int\left(\partial_{\theta} Y^*_{\ell0}\right)\left(\partial_{\theta}
Y_{\ell'0}\right)d \Omega, \\
 {\cal L}^{\pm 2}_1 {\cal A}_{\ell} &:=& \sum_{\ell'}{\cal A}_{\ell'}
        \int (\partial_{\theta} Y^*_{\ell0})\sin^2\theta (\partial_{\theta}
Y_{\ell'0}) d\Omega \nonumber \label{L1}\\
     &=& -\ell(\ell+3)Q_{\ell+2}Q_{\ell+1}{\cal A}_{\ell+2} + \left[ \ell^2
Q^2_{\ell+1} + (\ell+1)^2 Q^2_\ell\right] {\cal A}_{\ell}
     - (\ell-2)(\ell+1)Q_\ell Q_{\ell-1}{\cal A}_{\ell-2}, \\
 {\cal L}^{\pm 2}_2 {\cal A}_\ell &:=& \sum_{\ell'}{\cal A}_{\ell'}
        \int (\partial_{\theta} Y^*_{\ell 0})\sin\theta\cos\theta Y_{\ell' 0}
 d\Omega \nonumber \label{L2}\\
     &=& \ell Q_{\ell+1}Q_{\ell+2}{\cal A}_{\ell+2} + \left[\ell Q^2_{\ell+1}
 - (\ell +1)Q^2_\ell \right] {\cal A}_{\ell}
     - (\ell+1)Q_\ell Q_{\ell-1}{\cal A}_{\ell-2}, \\
 Q_\ell &:=& \sqrt{\frac{\ell^2}{(2\ell-1)(2\ell+1)}}.
\end{eqnarray}
From Equations (\ref{system}), (\ref{L1}) and (\ref{L2}) it is evident
that the presence of the dipole magnetic field causes the
eigenfunction of a specific torsional $(\ell,0)$ mode to be no longer
described by an angular part that corresponds to a single value of
$\ell$ (as in the nonmagnetized case). Instead, the eigenfunction of a
specific mode also acquires pieces with $\ell+2$, $\ell-2$ and higher
order dependence (through successive couplings). In order to solve the
full problem to a desired  accuracy, one has to solve a linear system of
equations of the type (\ref{system}) for required values of $\ell$,
up to some finite $\ell_{\rm max}$. Since the coupling is purely
a magnetic field effect, in the absence of a magnetic field the radial
functions ${\cal B}_\ell$, ${\cal C}_\ell$, and ${\cal D}_\ell$ vanish and the different
$\ell$ dependencies decouple.

\subsection{Neglecting $\ell \pm 2$ couplings}

As mentioned in the introduction, we limit our current study to
moderate magnetic field strengths of a few times $10^{15}$G, since
for larger values one would also have to take into account global
magnetosonic oscillations. We further simplify the numerical problem, 
by neglecting  the couplings to $\ell \pm 2$ terms in Equation
(\ref{system}). This is motivated by the expectation that higher-order
contributions will be less significant than the lowest-order piece of an
eigenfunction, since the higher-order pieces are sampling the
background magnetic field at smaller scales. This expectation may be
valid for the study of low-order modes. However, it is difficult to
predict the effect of $\ell \pm 2$ couplings on high-order modes
without solving the full two-dimensional problem. Our numerical 
results will thus be valid only to the extend that magnetic-field-induced
couplings to $\ell \pm 2$ terms can be neglected. 

Neglecting  $\ell \pm 2$ couplings leads to the eigenvalue equation
\begin{eqnarray}
 \Bigg[\mu + (1 + 2 \lambda_1)\frac{{a_1}^2}{\pi r^4}\Bigg]{\cal Y}''
    &+& \left\{\left(\frac{4}{r} + \Phi' - \Lambda'\right)\mu + \mu'
     + (1 + 2\lambda_1)\frac{a_1}{\pi r^4}\left[\left(\Phi' - \Lambda'\right)a_1
+ 2{a_1}'\right]\right\}{\cal Y}'\nonumber \\
    &+& \Bigg\{\left[\left(\epsilon + p + (1 +2\lambda_1)\frac{{a_1}^2}{\pi r^4}
\right)e^{2\Lambda}
     - \frac{\lambda_1 {{a_1}'}^2}{2\pi r^2}\right]\omega^2 e^{-2\Phi} \nonumber \\
    &&\hspace{1cm} - (\lambda-2)\left(\frac{ \mu e^{2\Lambda}}{r^2}
- \frac{\lambda_1{{a_1}'}^2}{2\pi r^4}\right)
     + (2 + 5\lambda_1)\frac{a_1}{2\pi r^4}\left\{\left(\Phi'
- \Lambda'\right){a_1}' + {a_1}''\right\}
     \Bigg\}{\cal Y} = 0, \label{system1}
\end{eqnarray}
where
\begin{eqnarray}
 \lambda &=& \ell(\ell+1), \\
 \lambda_1 &=& \ell Q_{\ell+1}^2 - (\ell+1) Q_{\ell}^2
=-\frac{\ell(\ell+1)}{(2\ell-1)(2\ell+3)}.
\end{eqnarray}
In order to solve a system of first-order ODEs, we define new
variables ${\cal Y}_1$ and ${\cal Y}_2$, through
\begin{eqnarray}
 {\cal Y}_1 &\equiv& {\cal Y} r^{1-\ell}, \label{Y1} \\
 {\cal Y}_2 &\equiv& \left[\mu + (1 +2 \lambda_1)
\frac{{a_1}^2}{\pi r^4}\right]e^{\Phi-\Lambda}{{\cal Y}}'r^{2-\ell}.
    \label{Y2}
\end{eqnarray}
In defining the new variables, we took into account the
form of the eigenfunction ${\cal Y}$ near the center,
where ${\cal Y}\sim r^{\ell-1}$. The final first-order system
of equations to be solved numerically for the real eigenvalues $\omega$
is then
\begin{eqnarray}
 {{\cal Y}_1}' &=& -\frac{\ell-1}{r}{\cal Y}_1
     + \frac{\pi r^3}{\pi r^4 \mu + (1 + 2\lambda_1) {a_1}^2}
e^{-\Phi + \Lambda}{\cal Y}_2,\label{eq:master3a} \\
 {{\cal Y}_2}' &=& -\Bigg[\left(\epsilon + p + (1 + 2\lambda_1)
\frac{{a_1}^2}{\pi r^4}
     - \lambda_1 e^{-2\Lambda}\frac{ {{a_1}'}^2}{2\pi r^2} \right)\omega^2
r e^{2(\Lambda-\Phi)}  \nonumber \\
    &&\hspace{1cm}- (\lambda-2)\left(\frac{ \mu e^{2\Lambda}}{r}
- \frac{\lambda_1{{a_1}'}^2}{2\pi r^3}\right)
     + (2+ 5\lambda_1) \frac{a_1 e^{2\Lambda}}{\pi r^3}
\left(\frac{a_1}{r^2} - 2\pi j_1\right)
       \Bigg]e^{\Phi - \Lambda}{\cal Y}_1
    - \frac{\ell+2}{r}{\cal Y}_2, \label{eq:master3b}
\end{eqnarray}
where Equation (\ref{Maxwell-J1}) was used in order to eliminate
the term of ${a_1}''$.

\subsection{Boundary conditions}

If one would solve the above system of equations from the center
of the star to the surface, then regularity at the center implies
the boundary condition
\begin{eqnarray}
 {\cal Y}_2 = (\ell-1)\left[\mu + (1 +2 \lambda_1)
\frac{{\alpha_c}^2}{\pi}\right]e^{\Phi}{{\cal Y}_1}\, .
\end{eqnarray}
If the equations (\ref{eq:master3a})-(\ref{eq:master3b}) are used for the
study of global oscillations, i.e. magnetosonic waves in the core and
the crust and shear waves in the crust, an appropriate boundary
condition at the base of the fluid/crust interface will be ${\cal Y}_2^{(-)}=
{\cal Y}_2^{(+)}$ which leads to the condition
\begin{eqnarray}
 {\cal Y}^{'(-)}= \left[1+\frac{1}{1+2\lambda_1}
\frac{u_s^2}{u_A^2}\right]{\cal Y}^{'(+)} .
\end{eqnarray}
This is analogous to the one used by \citet{GSA2006} in their study of
a corresponding toy problem.

However, here, we only consider torsional modes confined to the crust and
impose a zero traction condition at the base of the crust, which
implies that ${\cal Y}_2=0$ there. At the stellar surface, the
``zero-torque-at-surface" condition is imposed, i.e. $\delta
T^{(s)r}_{\ \ \phi}=0$ \citep{Schumaker1983}, which again implies
${\cal Y}_2=0$ with our choice of variables. We neglect the possible
presence of a thin fluid ocean.

\begin{figure}
\includegraphics[width=100mm]{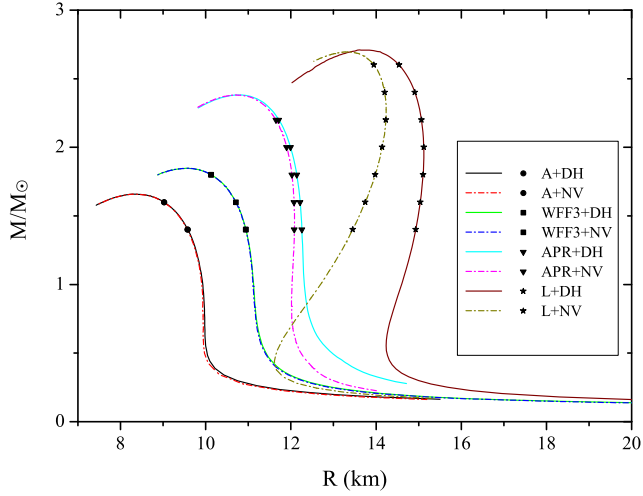}
\vspace{5mm}
 \caption{Mass-radius relationship of nonrotating equilibrium models
constructed with various high-density EoS, in combination with
the two choices for the crust EoS. Individual models for which
we compute torsional modes are shown with symbols.}
  \label{Fig:EoS}
\end{figure}

\section{Numerical Results}
\label{sec:VI}

\subsection{Description of the EoS and equilibrium models}

We have studied torsional modes for a variety of neutron star models
using four different equations of state for the core, ranging from a
very soft EoS (EoS A) \citep{EOS_A} to a very stiff (EoS L)
\citep{EOS_L}, with two intermediate ones, EoS WFF3 \citep{EOS_WFF3}
and APR \citep{EOS_APR}. For each EoS we constructed a number of
models, starting from a gravitational mass of $1.4 M_\odot$ and
reaching close to the maximum mass limit in increments of $0.2 M_
\odot$. In order to separately investigate the effect of the
composition of the crust, we matched the various high-density EoS to
two different proposed equations of state for the crust, one recent
derived by \citet{DH2001} (DH) and, for reference, and older EoS by
\citet{NV1973} (NV). Figure \ref{Fig:EoS} displays the mass-radius
relationship of nonrotating equilibrium models constructed with
the various high-density EoS, in combination with our two
choices for the crust EoS. The individual models for which we
compute torsional modes are shown with symbols, while their
detailed properties are listed in Table \ref{Tab:EOS_Models}.

From Figure \ref{Fig:EoS} it is evident that for the soft equations of
state (A and WFF3) the choice of the crust EoS does not affect
significantly the bulk properties of the star, while for the stiffest
EoS L models with the same mass have considerably different radii. The
reason for this is that the models constructed with the stiff EoS have
central densities that are much closer to the density of the base of
the crust than for soft EoS. The two crust EoS differ significantly
both in the detailed composition, as well as in the density at the
base of the crust, which is at $\rho\approx 2.4\times
10^{14}$gr/cm$^3$ for \citet{NV1973} and at $\rho\approx 1.28\times
10^{14}$gr/cm$^3$ for \citet{DH2001}. For the stiff EoS these
different properties of the crust have a considerable effect on the
bulk properties.


For the two crust EoS the shear modulus $\mu$ can be computed using
Equation (\ref{eq:mu}). For the NV EoS a simple fitting formula
was already derived by \citet{Duncan1998}
\begin{equation}
\mu= 1.267 \times 10^{30} \mbox{ erg cm}^{-3} \rho_{14}^{4/5},
\end{equation}
which we also adopt here. For the DH EoS, we derive the following
fitting (see also \citep{Piro2005})
\begin{eqnarray}
\mu&=& 1.74\times 10^{30} \mbox{ erg cm}^{-3} \rho_{14}^{4/3}
\left(\frac{Z}{53}\right)^2 \left(\frac{616}{A} \right)^{4/3}
\left(\frac{1-X_n}{0.41}\right)^{4/3}  \\
&=& 10^{30} \mbox{ erg cm}^{-3} \left(0.02123+0.37631 \rho_{14}
+3.13044 \rho_{14}^2-4.71841 \rho_{14}^3 + 2.46792 \rho_{14}^4\right)
\end{eqnarray}
where $X_n$ is the fraction of nucleons in the neutron gas outside nuclei.

\begin{table*}
 \centering
 \begin{minipage}{134mm}
\caption{ Mass ($M$), Radius ($R$) and relative crust thickness
    ($\Delta r/R$) for specific equilibrium models constructed
with various high-density EoS (A, WFF3, APR and L) in combination
with two crust EoS (DH and NV). The subscript in the model names is
indicative of the mass of each model.}
 \label{Tab:EOS_Models}
  \begin{tabular}{@{}lrrrclrrr@{}}
  \hline
   Model & $M/M_{\odot}$ & $R$(km) & $\Delta r/R$(\%) & |
 & Model & $M/M_{\odot}$ & $R$(km) & $\Delta r/R$(\%)   \\
 \hline
 A+DH$_{14}$    & 1.4&  9.49 & 5.15 & & A+NV$_{14}$   & 1.4 &  9.48 & 6.79 \\
 A+DH$_{16}$    & 1.6&  8.95 & 3.54 & & A+NV$_{16}$   & 1.6 &  8.94 & 4.69 \\
 WFF3+DH$_{14}$ & 1.4& 10.82 & 6.45 & & WFF3+NV$_{14}$& 1.4 & 10.82 & 8.53 \\
 WFF3+DH$_{16}$ & 1.6& 10.61 & 4.98 & & WFF3+NV$_{16}$& 1.6 & 10.61 & 6.54 \\
 WFF3+DH$_{18}$& 1.8& 10.03 & 3.54 & & WFF3+NV$_{18}$& 1.8 & 10.03 & 4.64 \\
 APR+DH$_{14}$  & 1.4& 12.10 & 7.72 & & APR+NV$_{14}$ & 1.4 & 11.93 & 9.98 \\
 APR+DH$_{16}$  & 1.6& 12.09 & 6.23 & & APR+NV$_{16}$ & 1.6 & 11.95 & 8.08 \\
 APR+DH$_{18}$  & 1.8& 12.03 & 5.05 & & APR+NV$_{18}$ & 1.8 & 11.92 & 6.54 \\
 APR+DH$_{20}$  & 2.0& 11.91 & 4.05 & & APR+NV$_{20}$ & 2.0 & 11.82 & 5.26 \\
 APR+DH$_{22}$  & 2.2& 11.65 & 3.15 & & APR+NV$_{22}$ & 2.2 & 11.59 & 4.11 \\
 L+DH$_{14}$    & 1.4& 14.66 & 10.16 & & L+NV$_{14}$   & 1.4 & 13.58 & 12.18 \\
 L+DH$_{16}$    & 1.6& 14.78 & 8.49 & & L+NV$_{16}$   & 1.6 & 13.82 & 10.22 \\
 L+DH$_{18}$    & 1.8& 14.83 & 7.11 & & L+NV$_{18}$   & 1.8 & 14.00 & 8.65 \\
 L+DH$_{20}$    & 2.0& 13.82 & 5.99 & & L+NV$_{20}$   & 2.0 & 14.09 & 7.32 \\
 L+DH$_{22}$    & 2.2& 14.73 & 5.02 & & L+NV$_{22}$   & 2.2 & 14.11 & 6.18 \\
 L+DH$_{24}$    & 2.4& 14.54 & 4.16 & & L+NV$_{24}$   & 2.4 & 14.02 & 5.15 \\
 L+DH$_{26}$    & 2.6& 14.13 & 3.32 & & L+NV$_{26}$   & 2.6 & 13.68 & 4.10 \\
\hline
\end{tabular}
\end{minipage}
\end{table*}

\subsection{Torsional-mode frequencies in the  non-magnetized limit}

Even though the excitation of these modes and the SGR activity are
likely related to the strong magnetic field, the torsional mode
frequencies are not affected by the magnetic field for $B<10^{15}$G
(at least when the latter is confined to the crust).
It is not clear whether magnetars have magnetic fields that
only approach this value or that are even stronger. One should
therefore first consider torsional mode frequencies in the
non-magnetized limit.

In Table \ref{Tab:tab1} we list the frequencies of the fundamental
and torsional modes $_\ell t_0$ for $\ell=2$ to $\ell=10$, for the
equilibrium models discussed earlier and shown in Table
\ref{Tab:EOS_Models}.
In addition to those models, we computed a few more i.e.
L+DH$_{17}$, L+NV$_{19}$, and L+NV$_{25}$ in an attempt to fit the 
observed frequencies to given models.
Among these models, the frequency of the fundamental $\ell=2$ mode
varies from 17 to 29 Hz i.e. depending on the stellar parameters it
can vary by up to 30-50\%. Limiting attention to models with
$M=1.4M_\odot$ only, the variation is from 21 to 29 Hz, which still
significant. Therefore, strong constraints of the high-density EoS
could be placed by observations of the fundamental $\ell=2$
torsional mode in magnetars, especially if the mass is deduced by
other means. On the other hand, the choice of the crust EoS does not
seem to affect significantly the frequency of the fundamental
torsional mode.  The typical variation observed for equal mass
models with different crust EoS is on the order of 1-5\%. The same
comments apply to the behavior of higher $\ell$ modes, which are
nearly equally spaced for $\ell \ge 3$, such that a scaling law can
be derived for each model.

The above picture is drastically altered when one considers the first
overtone, $_\ell t_1$, for which the obtained frequencies are shown in
Table \ref{Tab:tab2}.  We observe that our numerical data are in
agreement with the conclusion in \citet{HC1980} that the frequencies
are practically independent of the harmonic index $\ell$.
A second observation is that the variations in the frequencies due to
different choices of both the high-density and crust EoS are
significant.
Notice that these observations apply for the higher overtones ${}_{\ell}f_n$ as well.
The frequencies of the first overtones vary from 500-1200 Hz, while even for
models with the same high-density EoS and same mass the frequency
varies by up to 30\% when changing the crust EoS. This behavior is
quite useful given that there are suggestions for an observation in
the spectrum of SGR 1802+14 of a torsional mode with $n=1$ with a
frequency of 626.5Hz \citep{WS2006}
and at least one more at 1837 Hz.
Comparing this to our results in
Table \ref{Tab:tab2} one could immediately exclude the soft and
intermediate EoS A and WFF3, if the magnetic field is such that it
does not affect the torsional mode frequencies. The 1.4$M_\odot$ model
of EoS APR with an NV description of the crust agrees with the
observation, as does the 2.0$M_\odot$ model with EoS L and NV for the
crust. EoS L also seems to be compatible with the observation when
combined with the DH crust EoS for a mass somewhat larger than $1.6
M_\odot$.  Finally, one can observe also that as the stellar mass
increases the fundamental mode frequency decreases while the frequency
of the 1st overtone increases.
Note that this last observation also applies for the
frequencies of the higher overtones as well (also see Table \ref{Tab:tab3}).

In the tables we have marked the frequencies which are close to the
observed ones (within 2\%) as bold for the source SGR 1806-20,
underlined for the source SGR 1900+14 and italic for the source SGR
0526-66. One can observe that some of the stiff models fit quite
well to the observational data i.e. model L+NV$_{25}$
(where ${}_{17}f_0=$153 Hz) for SGR
1900+14. Still, model
A+DH$_{16}$ fits quite well with the observed frequencies of SGR
1900+14 and the $\ell=11$ fundamental mode (not shown in the
above tables) is exactly at 155 Hz.
 On the other hand, for SGR 1806-20 the models L+DH$_{17}$ (where ${}_{15}f_0=155$ Hz) or
L+NV$_{20}$ (where ${}_{15}f_0=152$ Hz) are in good agreement with the observational data.
According to our estimates, the observed frequencies of 18, 29, 92.5, 150, 626.5, 
and 1837 Hz would correspond to
${}_2f_0$, ${}_3f_0$, ${}_9f_0$, ${}_{15}f_0$, ${}_{\ell}f_1$,
and ${}_{\ell}f_4$, respectively,
although it is difficult to explain the observational data of $26$ Hz
because this data is very close to the other data of 29 Hz and
it may be difficult to explain both frequencies by crustal torsional modes only.
Additionally,  model APR+NV$_{14}$ also agrees well with the observational data for
SGR 1806-20. In the this non-magnetized limit, the currently available
observational data appear to exclude softer EoS, such as EoS A and
WFF3, if crustal torsional modes are invoked in interpreting the data.

\subsection{Implications for crust thickness}

As mentioned in the Introduction, one expects a simple relation
between the thickness of the crust, $\Delta r/R$ and the ratio of the
frequencies of the fundamental mode and higher harmonics, Equation
(\ref{eq:mode_spacing}). Using our numerical results, we can test the
validity of the expected relation and attempt to infer the thickness
of the crust from the observed frequencies in several SGRs. 
We construct the following empirical formula for the crust thickness 
\begin{equation}
\frac{\Delta r}{R} \approx {{}_\ell \beta_n}
 \frac{{}_\ell f_0}{{}_\ell f_n} \, .
 \label{eq:mode_spacing2}
\end{equation}
where ${_\ell \beta_n}$ are fitting coefficients.  Using our numerical
results presented in previous sections we find for the most
interesting case of $\ell=2$ and for $n=0$ and $n=1$ the fitted values
${_2\beta_1}=2.23\pm 0.29$ for all models with a crust described by
the NV EoS and ${_2 \beta_1}=2.19\pm 0.28$ for the corresponding
models with a crust described by the DH EoS. These values are somewhat
larger than the ones shown in the approximate relation
(\ref{eq:mode_spacing}), but this should be expected, since the
Equation (\ref{eq:mode_spacing}) was based on Newtonian dynamics and
heuristic arguments. 

Using the above empirical relation, important conclusions can be drawn
for the stellar models listed in Table 1. The observed frequencies of
SGR 1806-20 SGR, ${_\ell f_0}=18$Hz and ${_\ell f_1} =626.5$Hz suggest
a crust ratio $\Delta r/R \approx 0.06$ which combined with the
observed frequencies favors the model L+NV$_{22}$. Although,
this conclusion is based on a single observation, it is still very
important because it favors neutron star models with a very stiff
equation of state, implying considerably larger masses than the
typical ones. One should still keep in mind that there is an
uncertainty regarding the observed frequency at 18Hz. If the actual
frequency for the ${_\ell f_0}$ mode is at 29Hz then the empirical
formula (\ref{eq:mode_spacing2}) implies $\Delta r/R \approx 0.10$,
which suggest stellar models with considerable smaller mass such as
APR+NV$_{14}$ again, or ones with even smaller mass.

\begin{table*}
 \centering
 \begin{minipage}{140mm}
  \caption{Frequencies (in Hz) of the fundamental torsional modes in
the non-magnetized limit (i.e. $n=0$ and $B=0$). 
Values in bold type, underline, and italic type are within 2\% of
observational data for SGR 1806-20, SGR 1900+14, and SGR 0526-66,
respectively.}
  \label{Tab:tab1}
  \begin{tabular}{@{}lrrrrrrrrrr@{}}
  \hline
Model &   $\ell=2$ & $\ell=3$ & $\ell=4$ & $\ell=5$ &$\ell=6 $ & $\ell=7$ & $\ell=8$ & $\ell=9$& $\ell=10$ \\
 \hline
 A+DH$_{14}$    & {\underline{\bf 28.5}} & 45.1 & 60.5 & 75.4 & 90.1 & 104.7 & 119.2 & 133.7 & 148.1 \\
 A+DH$_{16}$    & 27.2 & {\em 43.0} & 57.7 & 72.0 & 86.0 &  99.9 & 113.8 & 127.6 & 141.3 \\
 WFF3+DH$_{14}$&  {\bf 26.3} & 41.6 & 55.9 & 69.7 & {\underline{83.3}} &  96.8 & 110.2 & 123.5 & 136.8 \\
 WFF3+DH$_{16}$ & 25.2 & 39.9 & {\underline{53.5}} & 66.8 & 79.8 &  {\bf 92.7} & 105.5 & 118.4 & 131.1 \\
 WFF3+DH$_{18}$ & 24.3 & 38.4 & 51.5 & 64.3 & 76.8 &  89.3 & 101.6 & 114.0 & 126.2 \\
 APR+DH$_{14}$  & 24.6 & 38.9 & 52.2 & 65.1 & 77.8 &  90.4 & 102.9 & 115.4 & 127.8 \\
 APR+DH$_{16}$  & 23.4 & 37.0 & 49.6 & 61.9 & 73.9 &  85.9 &  97.8 & 109.6 & 121.5 \\
 APR+DH$_{18}$  & 22.3 & 35.2 & 47.3 & 58.9 & 70.4 &  81.9 &  {\bf 93.2} & 104.5 & 115.7 \\
 APR+DH$_{20}$  & 21.3 & 33.6 & 45.1 & 56.3 & 67.3 &  78.1 &  89.0 &  99.7 & 110.5 \\
 APR+DH$_{22}$  & 20.2 & 31.9 & {\em 42.8} & {\underline{53.4}} & 63.8 &  74.1 &  {\underline{84.4}} &  94.6 & 104.8 \\
 L+DH$_{14}$    & 21.6 & 34.1 & 45.7 & 60.0 & 68.1 &  79.1 &  90.1 & 101.0 & 111.9 \\
 L+DH$_{16}$    & 20.6 & 32.5 & {\em 43.7} & {\underline{54.5}} & 65.1 &  75.6 &  86.1 &  96.5 & 106.9 \\
 L+DH$_{17}$    & 20.1 & 31.8 & 42.6 & {\underline{53.2}} & 63.5 &  73.8 &  {\underline{84.0}} &  {\bf 94.2} & 104.4 \\
 L+DH$_{18}$    & 19.7 & 31.1 & 41.7 & 52.0 & 62.2 &  72.2 &  82.3 &  {\bf 92.2} & 102.2 \\
 L+DH$_{20}$    & 18.9 & 29.9 & 40.2 & 50.1 & 59.9 &  69.6 &  79.2 &  88.8 & 98.4 \\
 L+DH$_{22}$    & {\bf 18.2} & {\bf 28.8} & 38.6 & 48.1 & 57.5 &  66.8 &  76.1 &  {\underline{85.3}} &  94.5 \\
 L+DH$_{24}$    & 17.5 & {\underline{27.7}} & 37.2 & 46.4 & 55.5 &  64.5 &  73.4 &  82.3 &  {\bf 91.2} \\
 L+DH$_{26}$    & 17.0 & 26.8 & 36.0 & 44.8 & {\underline{53.6}} &  62.3 &  70.9 &  79.5 &  88.1 \\
\hline
 A+NV$_{14}$    & {\bf 28.7} & 45.4 & 60.9 & 76.0 & {\bf 90.8} & 105.4 & 120.2 & 134.7 &149.2 \\
 A+NV$_{16}$    & 27.4 & {\em 43.3} & 58.1 & 72.4 & 86.6 & 100.6 & 114.5 & 128.4 &142.2 \\
 WFF3+NV$_{14}$ & 26.7 & 42.2 & 56.6 & 70.6 & {\underline{84.4}} &  98.1 & 111.6 & 125.2 &138.6 \\
 WFF3+NV$_{16}$ & 25.4 & 40.2 & {\underline{53.9}} & 67.2 & 80.4 &  {\bf 93.4} & 106.3 & 119.2 &132.0 \\
 WFF3+NV$_{18}$ & 24.4 & 38.6 & 51.7 & 64.5 & 77.1 &  89.6 & 102.0 & 114.4 &126.7 \\
 APR+NV$_{14}$  & 25.2 & 39.8 & {\underline{53.4}} & 66.6 & 79.5 &  {\bf 92.4} & 105.2 & 118.0 & 130.7 \\
 APR+NV$_{16}$  & 23.8 & 37.6 & 50.5 & 63.0 & 75.3 &  87.5 &  99.6 & 111.7 & 123.7 \\
 APR+NV$_{18}$  & 22.6 & 35.7 & 47.9 & 59.8 & 71.4 &  {\underline{83.0}} &  94.5 & 106.0 & 117.4 \\
 APR+NV$_{20}$  & 21.4 & 33.9 & 45.5 & 56.7 & 67.8 &  78.8 &  89.7 & 100.5 & 111.4 \\
 APR+NV$_{22}$  & 20.3 & 32.1 & {\em 43.1} & {\underline{53.8}} & 64.3 &  74.7 &  {\underline{85.0}} &  95.3 & 105.6 \\
 L+NV$_{14}$    & 23.2 & 36.6 & 49.2 & 61.3 & 73.3 &  {\underline{85.1}} &  96.9 & 108.6 & 120.3  \\
 L+NV$_{16}$    & 21.8 & 34.5 & 46.3 & 57.7 & 69.0 &  80.2 &  {\bf 91.3} & 102.4 & 113.4  \\
 L+NV$_{18}$    & 20.7 & 32.7 & {\em 43.9} & {\underline{54.7}} & 65.4 &  76.0 &  86.5 &  97.0 & 107.4 \\
 L+NV$_{19}$    & 20.2 & 31.9 & {\em 42.8} & {\underline{53.3}} & 63.8 &  74.1 &  {\underline{84.3}} &  94.6 & 104.8 \\
 L+NV$_{20}$    & 19.7 & 31.1 & 41.7 & 52.1 & 62.2 &  72.3 &  82.3 &  {\bf 92.3} & 102.2 \\
 L+NV$_{22}$    & 18.8 & 29.7 & 39.8 & 49.7 & 59.4 & 69.0 & 78.6 & 88.1 & 97.6 \\
 L+NV$_{24}$    & {\bf 18.0} & {\underline{28.4}} & 38.1 & 47.5 & 56.8 & 65.9 & 75.1 & {\underline{84.2}} & {\bf 93.3} \\
 L+NV$_{25}$    & 17.6 & {\underline{27.8}} & 37.3 & 46.5 & 55.5 & 64.5 & 73.5 & {\underline{82.4}} & {\bf 91.2} \\
 L+NV$_{26}$    & 17.2 & 27.2 & 36.5 & 45.5 & {\underline{54.4}} & 63.2 & 71.9 & 80.6 & 89.3  \\
\hline
\end{tabular}
\end{minipage}
\end{table*}

\begin{table*}
 \centering
 \begin{minipage}{130mm}
  \caption{Frequencies (in Hz) of the first overtone of torsional
    modes in the non-magnetized limit (i.e. $n=1$ and $B=0$).
    Values in
    bold type are within about 2\% of observational data for SGR 1806-20.}
  \label{Tab:tab2}
  \begin{tabular}{@{}lrrrrrrrrrr@{}}
  \hline
   EoS & $\ell=2$ & $\ell=3$ & $\ell=4$ & $\ell=5$ &$\ell=6 $ & $\ell=7$ & $\ell=8$ & $\ell=9$& $\ell=10$ \\
 \hline
 A+DH$_{14}$    & 1206.2 & 1206.8 & 1207.6 & 1208.6 & 1209.7 & 1211.1 & 1212.7 & 1214.5 & 1216.4 \\
 A+DH$_{16}$    & 1531.3 & 1531.7 & 1532.3 & 1533.0 & 1533.9 & 1534.9 & 1536.0 & 1537.3 & 1538.7 \\
 WFF3+DH$_{14}$ &  942.4 &  943.0 &  943.9 &  945.0 &  946.3 &  947.8 &  949.6 &  951.5 &  953.7 \\
 WFF3+DH$_{16}$ & 1101.1 & 1101.6 & 1102.3 & 1103.2 & 1104.2 & 1105.4 & 1106.8 & 1108.3 & 1110.0 \\
 WFF3+DH$_{18}$ & 1367.2 & 1367.6 & 1368.1 & 1368.7 & 1369.5 & 1370.4 & 1371.4 & 1372.5 & 1373.8 \\
 APR+DH$_{14}$  &  761.3 &  762.0 &  762.9 &  764.1 &  765.5 &  767.1 &  769.0 &  771.1 &  773.4 \\
 APR+DH$_{16}$  &  860.2 &  860.8 &  861.5 &  862.5 &  863.6 &  864.9 &  866.4 &  868.1 &  869.9 \\
 APR+DH$_{18}$  &  965.6 &  966.0 &  966.6 &  967.4 &  968.3 &  969.4 &  970.6 &  972.0 &  973.5 \\
 APR+DH$_{20}$  & 1083.3 & 1083.6 & 1084.1 & 1084.8 & 1085.5 & 1086.4 & 1087.3 & 1088.4 & 1089.7 \\
 APR+DH$_{22}$  & 1238.0 & 1238.3 & 1238.7 & 1239.2 & 1239.8 & 1240.5 & 1241.2 & 1242.1 & 1243.1 \\
 L+DH$_{14}$    &  530.2 &  531.0 &  532.0 &  533.3 &  534.8 &  536.6 &  538.7 &  540.9 &  543.4 \\
 L+DH$_{16}$    &  586.4 &  587.0 &  587.9 &  588.9 &  590.2 &  591.7 &  593.4 &  595.2 &  597.3 \\
 L+DH$_{17}$    &  {\bf 617.0} &  {\bf 617.6} &  {\bf 618.4} &  {\bf 619.4} &  {\bf 620.5} &  {\bf 621.9} &  {\bf 623.4} &  {\bf 625.1} &  {\bf 627.0} \\
 L+DH$_{18}$    &  648.1 &  648.6 &  649.3 &  650.2 &  651.3 &  652.5 &  653.9 &  655.5 &  657.2 \\
 L+DH$_{20}$    &  712.5 &  712.9 &  713.5 &  714.2 &  715.1 &  716.2 &  717.3 &  718.7 &  720.1 \\
 L+DH$_{22}$    &  788.1 &  788.5 &  789.0 &  789.6 &  790.3 &  791.2 &  792.2 &  793.3 &  794.5 \\
 L+DH$_{24}$    &  874.2 &  874.5 &  875.0 &  875.5 &  876.1 &  876.8 &  877.7 &  878.6 &  879.6 \\
 L+DH$_{26}$    &  995.1 &  995.4 &  995.7 &  996.2 &  996.7 &  997.3 &  997.9 &  998.7 &  999.5 \\
\hline
 A+NV$_{14}$    &  951.0 &  951.7 &  952.8 &  954.0 &  955.6 &  957.3 &  959.4 &  961.6 &  964.2 \\
 A+NV$_{16}$    & 1190.8 & 1191.4 & 1192.1 & 1193.1 & 1194.2 & 1195.5 & 1197.0 & 1198.6 & 1200.5 \\
 WFF3+NV$_{14}$ &  740.8 &  741.6 &  742.7 &  744.1 &  745.8 &  747.8 &  750.0 &  752.5 &  755.3 \\
 WFF3+NV$_{16}$ &  865.8 &  866.5 &  867.4 &  868.5 &  869.8 &  871.3 &  873.1 &  875.0 &  877.2 \\
 WFF3+NV$_{18}$ & 1069.8 & 1070.3 & 1071.0 & 1071.8 & 1072.8 & 1073.9 & 1075.2 & 1076.7 & 1078.4 \\
 APR+NV$_{14}$  &  {\bf 616.5} &  {\bf 617.3} &  {\bf 618.5} &  {\bf 620.0} &  {\bf 621.8} &  {\bf 623.9} &  {\bf 626.3} &  {\bf 628.9} &  {\bf 631.8} \\
 APR+NV$_{16}$  &  688.5 &  689.2 &  690.2 &  691.4 &  692.8 &  694.5 &  696.4 &  698.5 &  700.9 \\
 APR+NV$_{18}$  &  769.4 &  770.0 &  770.7 &  771.7 &  772.9 &  774.2 &  775.8 &  777.5 &  779.5 \\
 APR+NV$_{20}$  &  858.3 &  858.7 &  859.4 &  860.2 &  861.1 &  862.2 &  863.5 &  864.9 &  866.4 \\
 APR+NV$_{22}$  &  974.4 &  974.7 &  975.2 &  975.9 &  976.6 &  977.5 &  978.5 &  979.6 &  980.9 \\
 L+NV$_{14}$    &  483.6 &  484.6 &  485.9 &  487.5 &  489.4 &  491.6 &  494.1 &  496.9 &  500.1 \\
 L+NV$_{16}$    &  524.6 &  525.4 &  526.4 &  527.7 &  529.3 &  531.1 &  533.2 &  535.6 &  538.1 \\
 L+NV$_{18}$    &  567.7 &  568.3 &  569.2 &  570.3 &  571.6 &  573.2 &  574.9 &  576.9 &  579.0 \\
 L+NV$_{19}$    &  589.7 &  590.3 &  591.1 &  592.1 &  593.4 &  594.8 &  596.4 &  598.2 &  600.1 \\
 L+NV$_{20}$    &  {\bf 614.9} &  {\bf 615.4} &  {\bf 616.2} &  {\bf 617.1} &  {\bf 618.2} &  {\bf 619.5} &  {\bf 621.0} &  {\bf 622.6} &  {\bf 624.4} \\
 L+NV$_{22}$    &  667.1 &  667.5 &  668.2 &  668.9 &  669.9 &  670.9 &  672.2 &  673.6 &  675.1 \\
 L+NV$_{24}$    &  729.1 &  729.5 &  730.0 &  730.7 &  731.4 &  732.3 &  733.4 &  735.6 &  735.9 \\
 L+NV$_{25}$    &  769.8 &  770.2 &  770.7 &  771.3 &  772.0 &  772.8 &  773.8 &  774.8 &  776.0 \\
 L+NV$_{26}$    &  824.1 &  824.4 &  824.8 &  825.3 &  826.0 &  826.7 &  827.6 &  828.5 &  829.6 \\
\hline
\end{tabular}
\end{minipage}
\end{table*}

\begin{table*}
 \centering
 \begin{minipage}{130mm}
  \caption{Frequencies (in Hz) of the overtone of torsional
    modes for $\ell=2$ and $n=2$, 3, and 4 in the non-magnetized limit (i.e. $B=0$).
    Values in
    bold type are within about 2\% of observational data for SGR 1806-20.}
  \label{Tab:tab3}
  \begin{tabular}{@{}lrrr@{}}
  \hline
   EoS & $n=2$ & $n=3$ & $n=4$ \\
 \hline
 A+DH$_{14}$    & 2011.4 & 2774.8 & 3476.4 \\
 A+DH$_{16}$    & 2552.6 & 3521.8 & 4412.2 \\
 WFF3+DH$_{14}$ & 1572.0 & 2164.7 & 2708.7 \\
 WFF3+DH$_{16}$ & {\bf 1835.8} & 2530.7 & 3167.6 \\
 WFF3+DH$_{18}$ & 2278.2 & 3142.1 & 3933.3 \\
 APR+DH$_{14}$  & 1270.0 & 1747.9 & 2186.8 \\
 APR+DH$_{16}$  & 1434.7 & 1976.0 & 2472.5 \\
 APR+DH$_{18}$  & 1610.0 & 2218.8 & 2776.6 \\
 APR+DH$_{20}$  & {\bf 1805.4} & 2489.4 & 3115.8 \\
 APR+DH$_{22}$  & 2062.9 & 2844.2 & 3559.7 \\
 L+DH$_{14}$    &  885.1 & 1217.6 & 1525.1 \\
 L+DH$_{16}$    &  978.6 & 1347.9 & 1688.6 \\
 L+DH$_{17}$    & 1029.7 & 1418.7 & 1777.3 \\
 L+DH$_{18}$    & 1081.5 & 1490.4 & {\bf 1867.3} \\
 L+DH$_{20}$    & 1188.4 & 1639.2 & 2054.1 \\
 L+DH$_{22}$    & 1314.4 & {\bf 1813.3} & 2272.4 \\
 L+DH$_{24}$    & 1457.6 & 2011.5 & 2521.1 \\
 L+DH$_{26}$    & 1658.5 & 2289.4 & 2869.5 \\
\hline
 A+NV$_{14}$    & 1687.9 & 2390.3 & 2905.7 \\
 A+NV$_{16}$    & 2113.5 & 2997.8 & 3675.8 \\
 WFF3+NV$_{14}$ & 1315.1 & {\bf 1860.7} & 2250.9 \\
 WFF3+NV$_{16}$ & 1536.7 & 2176.5 & 2647.9 \\
 WFF3+NV$_{18}$ & 1898.3 & 2691.5 & 3295.1 \\
 APR+NV$_{14}$  & 1094.5 & 1547.3 & {\bf 1864.0} \\
 APR+NV$_{16}$  & 1222.4 & 1730.5 & 2098.8 \\
 APR+NV$_{18}$  & 1365.6 & 1934.5 & 2355.3 \\
 APR+NV$_{20}$  & 1523.2 & 2159.5 & 2641.2 \\
 APR+NV$_{22}$  & 1728.8 & 2452.0 & 3008.3 \\
 L+NV$_{14}$    &  858.9 & 1212.6 & 1453.1 \\
 L+NV$_{16}$    &  931.5 & 1316.9 & 1586.8 \\
 L+NV$_{18}$    & 1007.9 & 1426.4 & 1727.1 \\
 L+NV$_{19}$    & 1047.1 & 1482.6 & 1799.9 \\
 L+NV$_{20}$    & 1091.6 & 1546.0 & 1879.5 \\
 L+NV$_{22}$    & 1184.1 & 1678.3 & 2048.6 \\
 L+NV$_{24}$    & 1294.1 & {\bf 1835.3} & 2248.6 \\
 L+NV$_{25}$    & 1366.2 & 1937.6 & 2375.4 \\
 L+NV$_{26}$    & 1462.3 & 2074.3 & 2546.7 \\
\hline
\end{tabular}
\end{minipage}
\end{table*}

\subsection{Effect of the magnetic field}

We have already discussed earlier the possible effect of the magnetic field
on the frequencies of the various torsional modes, when the magnetic
field is confined to the crust. An approximate
formula for the effect of the magnetic field based on Newtonian
physics has been suggested by \citet{Duncan1998}, see also a
detailed discussion by \citet{Messios2001}. The formula suggests
that the corrections in the frequency coming from the magnetic field
scale as $B^2$ i.e.
\begin{equation}
\frac{{}_\ell f_n}{{}_\ell f_n^{(0)}}= \left[ 1+
\left(\frac{B}{B_\mu}\right)^2\right]^{1/2} \label{eq:magnet1}
\end{equation}
where ${}_\ell f_n^{(0)}$ is the frequency in the absence of any
magnetic field and $B_\mu$ is a typical magnetic field strength at
which magnetic field effects on torsional modes have become important,
which we take here to be $B_\mu=4\times 10^{15}$ G. In the limit of
$B<<B_\mu$, this result agrees with earlier studies of the effect of
the magnetic field on spheroidal modes derived by \citet{Carroll1986}
and \citet{Unno1989}, where the effect of the magnetic field has been
shown to have the behavior
\begin{equation}
f\approx f^{(0)}+ {}_\ell {\tilde \alpha}_n B^2 \label{eq:magnet2}
\end{equation}
where ${}_\ell {\tilde \alpha}_n$ is a coefficient depending on the
parameters of the star ($M$, $R$ and EoS) and it can be derived
easily if the eigenfunctions of a specific mode are available.
When one uses general relativity to calculate the frequencies of the
torsional modes the relativistic effects such as the relativistic
form of the sound speed, the redshift corrections and the EoS affect
the weighting factor ${}_\ell \alpha_n$ of equation (\ref{eq:magnet2})
and have to be taken properly into account.

In our numerical calculations of the torsional modes of magnetized
neutron stars for the various EoS mentioned earlier we derived lists
of frequencies for every model, for magnetic field strengths up to $B=
10^{17}$ G. By numerical fitting we have thus found the
coefficients $_\ell \alpha_n$ of the following empirical formula
\begin{equation}
\frac{{}_\ell f_n}{{}_\ell f_n^{(0)}}\approx \left[ 1+ {}_\ell\alpha_n
\left(\frac{B}{B_\mu}\right)^2 \right]^{1/2}\, . \label{eq:fit_l2}
\end{equation}

The coefficients ${}_\ell\alpha_n$ are listed in Table
\ref{Tab:fit_factors} and the way that the various EoS (both for the
core and the crust) affect the torsional frequencies becomes apparent.
In general, it seems that models constructed with the DH equation of
state are affected significantly more by the magnetic field than
models following the NV EoS, for all harmonics. As a result, the
large differences in the torsional mode frequencies between models
constructed with the stiffest EoS L but with different crust EoS
are diminished by magnetic field effects around $B=4\times 10^{15}$G
and reversed for larger values of $B$.

When the
magnetic field strength is equal to $B=B_\mu=4\times10^{15}$G the
frequencies have increased by up to 35\% for the fundamental torsional
mode (independently of the value of $\ell$) and up to 100\% for
the first overtone. The results suggest that for magnetic field
strengths exceeding roughly 10$^{15}$ G the shift in the
frequencies is significant and should be taken into account in any
attempt to fit the observational data with specific stellar models.

\begin{figure}
\includegraphics[width=100mm]{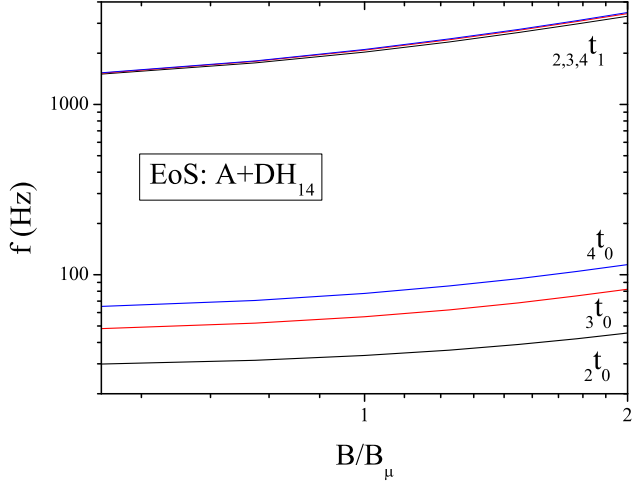}
 \caption{The frequencies of the fundamental $n=0$ and the first overtone for
 $\ell=2, 3$ and 4 torsional
 modes as functions of the normalized magnetic field ($B/B_\mu$).
 The neutron star mass is 1.4$M_\odot$ and we show results for
 EoS A+DH.  }
  \label{Fig:A_DH_n01_l234}
\end{figure}

\begin{figure}
\includegraphics[width=100mm]{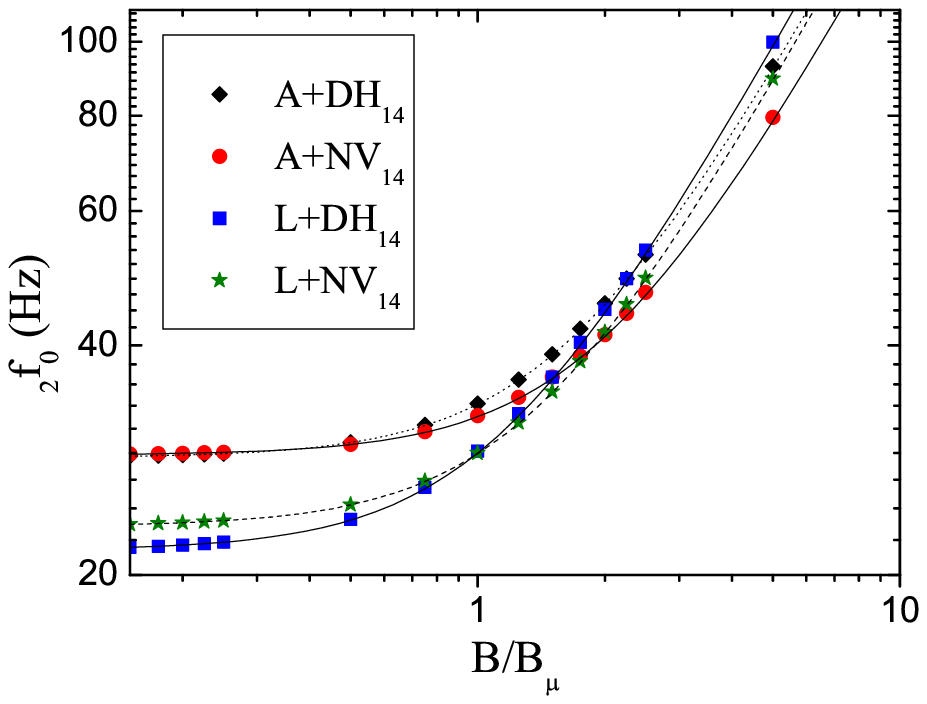}
 \caption{The frequencies of the fundamental $n=0$ and $\ell=2$ torsional
   mode as functions of the normalized magnetic field ($B/B_\mu$).
   The neutron star masses are 1.4$M_\odot$ and we show only results
   for four EoS i.e. A+DH, A+NV, L+DH and L+NV. The lines correspond
   to fits according to the empirical formula, Eq. (\ref{eq:fit_l2}), with
   coefficient values from Table \ref{Tab:fit_factors}.
   As seen here, our empirical formula, Eq. (79), agrees very well 
   with the numerical results.  We stress that these results were
   obtained in the approximation of neglecting magnetic-field-induced 
   deformations of the background star and couplings to $\ell \pm 2$
   terms.}
  \label{Fig:l2_B_eosAL}
\end{figure}

In Figures \ref{Fig:A_DH_n01_l234} and \ref{Fig:l2_B_eosAL} we show
some examples of the effect of the magnetic field on the torsional mode
frequencies. Notice that for $B<B_\mu$ the effect of the magnetic
field on the frequencies follows a quadratic increase, in agreement
with the approximate relation (\ref{eq:magnet2}). On the other hand,
when $B>B_\mu$ the modes change character and become dominated by
the magnetic field, while the frequencies tend to become less 
sensitive to the stellar parameters. In this regime, additional
effects, such as the coupling to global magnetosonic waves and to
higher-order harmonics should be taken into account.

\begin{table*}
 \centering
 \begin{minipage}{80mm}
  \caption{The values for the fitting factors ${}_\ell\alpha_n$
  of equation  (\ref{eq:fit_l2}).
  The fitting factors have been calculated for magnetic field strength up to $10^{17}$ G.}
\label{Tab:fit_factors}
  \begin{tabular}{@{}lrrrrrrr@{}}
  \hline
   Model  & $_2\alpha_0$ &  $_2\alpha_1$ &$_5\alpha_0$ &  $_5\alpha_1$
    & $_8\alpha_0$ &  $_8\alpha_1$  \\
 \hline
 A+DH$_{14}$    & 0.38 & 1.44 & 0.67 & 1.63 & 0.70 & 1.65\\
 A+DH$_{16}$    & 0.39 & 1.34 & 0.70 & 1.52 & 0.73 & 1.55 \\
 WFF3+DH$_{14}$ & 0.45 & 1.55 & 0.70 & 1.75 & 0.72 & 1.77 \\
 WFF3+DH$_{16}$ & 0.42 & 1.43 & 0.68 & 1.62 & 0.71 & 1.64 \\
 APR+DH$_{14}$  & 0.54 & 1.63 & 0.71 & 1.84 & 0.72 & 1.87 \\
 APR+DH$_{16}$  & 0.50 & 1.53 & 0.70 & 1.73 & 0.72 & 1.75 \\
 APR+DH$_{18}$  & 0.48 & 1.44 & 0.70 & 1.63 & 0.72 & 1.66 \\
 APR+DH$_{20}$  & 0.48 & 1.37 & 0.70 & 1.55 & 0.72 & 1.57 \\
 APR+DH$_{22}$  & 0.47 & 1.32 & 0.72 & 1.49 & 0.75 & 1.51 \\
 L+DH$_{14}$    & 0.80 & 1.83 & 0.77 & 2.06 & 0.75 & 2.09 \\
 L+DH$_{16}$    & 0.73 & 1.68 & 0.73 & 1.90 & 0.73 & 1.93 \\
 L+DH$_{18}$    & 0.67 & 1.59 & 0.72 & 1.80 & 0.72 & 1.83 \\
 L+DH$_{20}$    & 0.63 & 1.50 & 0.71 & 1.70 & 0.71 & 1.73 \\
 L+DH$_{22}$    & 0.60 & 1.44 & 0.71 & 1.64 & 0.72 & 1.66 \\
 L+DH$_{24}$    & 0.57 & 1.38 & 0.71 & 1.56 & 0.72 & 1.59 \\
 L+DH$_{26}$    & 0.55 & 1.32 & 0.72 & 1.50 & 0.74 & 1.53 \\
\hline
 A+NV$_{14}$    &  0.26 & 0.56 & 0.32 & 0.63 & 0.33 & 0.64 \\
 A+NV$_{16}$    &  0.24 & 0.50 & 0.32 & 0.57 & 0.33 & 0.58 \\
 WFF3+NV$_{14}$ & 0.34 & 0.60 & 0.33 & 0.68 & 0.33 & 0.69 \\
 WFF3+NV$_{16}$ & 0.30 & 0.55 & 0.33 & 0.62 & 0.33 & 0.63 \\
 APR+NV$_{14}$  & 0.42 & 0.63 & 0.35 & 0.72 & 0.33 & 0.73 \\
 APR+NV$_{16}$  & 0.38 & 0.58 & 0.34 & 0.66 & 0.33 & 0.67 \\
 APR+NV$_{18}$  & 0.35 & 0.55 & 0.33 & 0.62 & 0.33 & 0.63 \\
 APR+NV$_{20}$  & 0.32 & 0.52 & 0.33 & 0.59 & 0.33 & 0.60 \\
 APR+NV$_{22}$  & 0.30 & 0.49 & 0.34 & 0.56 & 0.34 & 0.57 \\
 L+NV$_{14}$    & 0.55 & 0.69 & 0.37 & 0.78 & 0.34 & 0.79 \\
 L+NV$_{16}$    & 0.51 & 0.64 & 0.36 & 0.73 & 0.33 & 0.73 \\
 L+NV$_{18}$    & 0.47 & 0.60 & 0.35 & 0.68 & 0.33 & 0.69 \\
 L+NV$_{20}$    & 0.44 & 0.57 & 0.34 & 0.64 & 0.33 & 0.65 \\
 L+NV$_{22}$    & 0.41 & 0.54 & 0.34 & 0.61 & 0.33 & 0.62 \\
 L+NV$_{24}$    & 0.39 & 0.51 & 0.34 & 0.58 & 0.33 & 0.59 \\
 L+NV$_{25}$    & 0.38 & 0.50 & 0.34 & 0.57 & 0.34 & 0.58 \\
 L+NV$_{26}$    & 0.36 & 0.49 & 0.35 & 0.56 & 0.34 & 0.57  \\
\hline
\end{tabular}
\end{minipage}
\end{table*}

\section{Summary and Discussion}

We have derived the formalism for computing torsional oscillations of
relativistic stars endowed with a strong dipole magnetic field,
confined to the crust. Our equations are valid in the Cowling
approximation (no spacetime perturbations) and we have neglected the
effect of the magnetic field on the equilibrium configuration and on
the coupling of torsional modes to $\ell \pm 2$ terms and to
global magnetosonic modes. Under these approximations, our formalism
allows us to obtain an estimate of the magnetic field effects on
torsional modes, up to moderate values of the magnetic field strength
(up to a few times $B_\mu=4 \times 10^{15}$G).  These moderate values
of the magnetic field strength are appropriate for models of
magnetars, for which there is strong evidence that they are at the
heart of the SGR phenomenon.

We have done a systematic search of parameter space by computing
torsional mode frequencies for various values of the harmonic index
$\ell$ and for various overtones, using an extended sample of models
of compact stars. These models vary in mass, high-density equation of
state and crust model, uniformly covering the allowed mass vs. radius
parameter space.  Our numerical results have shown that torsional mode
frequencies are sensitive to the crust model if the high-density
equation of state is very stiff (such as EoS L). In addition, torsional mode
frequencies are drastically affected by a dipole magnetic field, if
the latter has a strength exceeding roughly $10^{15}$G. The effect
of the magnetic field is surprisingly sensitive to the adopted crust
model. Using our extended numerical results we have derived empirical
relations for the effect of the magnetic field on torsional modes as
well as for the crust thickness.  We compare our numerical results to
observed frequencies in SGRs and find that certain high-density EoS
and mass values are favored over others in the non-magnetized limit.
On the other hand, if the magnetic field is strong, then its effect
has to be taken into account in attempts to formulate a theory of
asteroseismology for magnetars. This topic, as well as the inclusion
of global magnetosonic modes will be discussed in a separate
publication \citep{SKSV06}.

\section*{Acknowledgments}

We are grateful to Nils Andersson, Demetrios Papadopoulos, Adamantios
Stavridis and Miltos Vavoulidis for useful discussions.  This work was
supported by the Marie-Curie grant MIF1-CT-2005-021979 and the
Pythagoras II program of GSRT.


\appendix
\section[]{Monopole magnetic field}
\label{sec:IV}

If the magnetic field is assumed to be axisymmetric,
Maxwell's equations (\ref{Maxwell0-1}) become
\begin{equation}
 H^r_{\ ,r} + H^{\theta}_{\ ,\theta} + \left(\Lambda' + \frac{2}{r}\right)
H^r + \cot\theta H^{\theta}=0.
    \label{Maxwell-b}
\end{equation}
As a toy model, we consider a monopolar magnetic field in the crust
\begin{eqnarray}
 H^r &=& H^r(r) \neq 0, \\
 H^{\theta} &=& 0.
\end{eqnarray}
For this case, the solution of Maxwell's equation (\ref{Maxwell-b}) we
is \citep{Messios2001}
\begin{equation}
 H^r = \frac{1}{r^2}e^{-\Lambda} C,
\end{equation}
where $C$ is some constant.  For this case the perturbation equation
(\ref{master-equation}) is rewritten as
\begin{eqnarray}
 -(\epsilon + p + H^r H_r)\omega^2 e^{-2\Phi+2\Lambda}{\cal Y} &=& \left(\mu
+ H^r H_r\right) {\cal Y}'' \nonumber \\
     &+& \left[\left(\frac{4}{r} + \Phi' - \Lambda' \right)\mu + \mu'
     + (\Phi' - \Lambda')H^r H_r\right]{\cal Y}'
     - \frac{(\ell+2)(\ell-1)}{r^2}\mu e^{2\Lambda} {\cal Y}
\label{perturbation-equation-s}
\end{eqnarray}
If we introduce the new variables, such as
\begin{eqnarray}
 {\cal Y}_1 &\equiv& {\cal Y}, \\
 {\cal Y}_2 &\equiv& \left(\mu + H^r H_r\right)
e^{\Phi - \Lambda}{{\cal Y}_1}',
\end{eqnarray}
then equations (\ref{perturbation-equation-s}) reduce
\begin{eqnarray}
 {{\cal Y}_1}' &=& \frac{1}{\mu + H^r H_r}e^{-\Phi + \Lambda}{\cal Y}_2,\\
 {{\cal Y}_2}' &=& \left[\frac{(\ell+2)(\ell-1)}{r^2}\mu e^{2\Phi}
       - (\epsilon + p + H^r H_r)\omega^2 \right]e^{-\Phi+\Lambda} {\cal Y}_1
       - \frac{4}{r} {\cal Y}_2
\end{eqnarray}

\begin{figure}
\includegraphics[width=100mm]{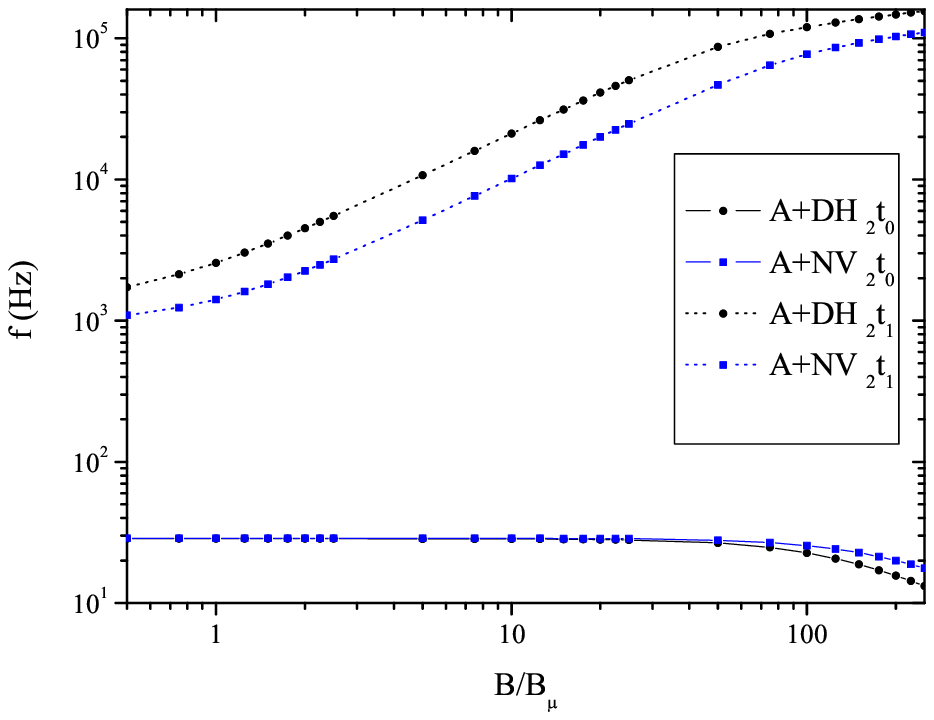}
 \caption{The frequencies of the $_2t_0$ and $_2t_1$ torsional
   modes as functions of the normalized magnetic field ($B/B_\mu$) for
   EoS A+DH and A+NV are shown.  The magnetic field has monopole
   geometry while the neutron stars have masses 1.4$M_\odot$. }
  \label{Fig:radial}
\end{figure}

In Figure \ref{Fig:radial} we show the fundamental and first overtone
frequencies for neutron star models with mass 1.4$M_\odot$. It is
apparent that in the case of a monopole magnetic field the frequencies
(especially of the fundamental mode) will be marginally affected by
the magnetic field and only for stengths on the order of 10$^{17}$
G its influence becomes significant. This is in qualitative
agreement with the conclusions in \cite{Messios2001}. In addition, 
we confirm the finding in \cite{Messios2001} that the frequency
of the fundamental mode decreases for a monopole magnetic field, as 
is apparent in Figure \ref{Fig:radial}.


\begin{thebibliography}{99}


\bibitem[\protect\citeauthoryear{Akmal et al.}{1998}]{EOS_APR}
    Akmal A., Pandharipande V.R., Ravenhall D.G., 1998, Phys. Rev.
    C 58, 1804

\bibitem[\protect\citeauthoryear{Barat et al.}{1983}]{Barat1983}
    Barat et al., 1983,  A\&A, 126, 400

\bibitem[\protect\citeauthoryear{Bocquet et al.}{1995}]{Bocquet1995}
    Bocquet M., Bonazzola S., Gourgoulhon E., Novak J., 1995,  A\&A, 301, 757

\bibitem[\protect\citeauthoryear{Bonazzola et al.}{1993}]{Bonazzola1993}
    Bonazzola S., Gourgoulhon E., Salgado M., Marck J.A., 1993, A\&A, 278, 421

\bibitem[\protect\citeauthoryear{Braithwaite \& Spruit}{2006}]{BS2006}
   Braithwaite J., Spruit H.C., A\&A, 450, 1097

\bibitem[\protect\citeauthoryear{Carroll et al.}{1986}]{Carroll1986}
   Carroll B.W., Zweibel E.G., Hansen C.., McDermot P.N., Savedoff M.P.,
   Thomas J.H., Van Horn H.M., 1986, ApJ, 305, 767

\bibitem[\protect\citeauthoryear{Carter \& Quintana}{1972}]{CQ1972}
   Carter B., Quintana H., 1973,
   Proc. R. Soc. Lond., A331, 57

\bibitem[\protect\citeauthoryear{Douchin \& Haensel}{2001}]{DH2001}
   Douchin F., Haensel P.,  2001, A\& A, 380, 151

\bibitem[\protect\citeauthoryear{Duncan}{1998}]{Duncan1998}
    Duncan R.C.,1998, ApJ, 1998,  498, L45

\bibitem[\protect\citeauthoryear{Duncan \& Thompson}{1992}]{DT1992}
    Duncan R.C., Thompson C., 1992, ApJ, 392, L9

\bibitem[\protect\citeauthoryear{Glampedakis et al.}{2006}]{GSA2006}
    Glampedakis K., Samuelson L., Andersson N., 2006, astro-ph/0605461

\bibitem[\protect\citeauthoryear{Hansen \& Cioffi}{1980}]{HC1980}
   Hansen C., Cioffi D.F, 1980, ApJ, 238, 740

\bibitem[\protect\citeauthoryear{Hurley et al.}{1999}]{Hurley1999}
    Hurley K. et al., 1999, Nature, 397, 41

\bibitem[\protect\citeauthoryear{Israel et al.}{2005}]{Israel2005}
    Israel G. et al., 2005, ApJ, 628, L53

\bibitem[\protect\citeauthoryear{Konno et al.}{1999}]{Konno1999}
    Konno K., Obata T., Kojima Y., 1999, A\&A, 352, 211

\bibitem[\protect\citeauthoryear{Lee}{2006}]{Lee2006}
    Lee, U., 2006, astro-ph/0610182

\bibitem[\protect\citeauthoryear{Leins}{1994}]{Leins1994}
    Leins M., 1994, PhD Thesis, University of T\"ubingen

\bibitem[\protect\citeauthoryear{Mazets et al.}{1979}]{Mazets1979}
    Mazets E.P. et al.,  1979, Nature 282, 587

\bibitem[\protect\citeauthoryear{McDermott et al.}{1988}]{McDermott1988}
   McDermott P.N., Van Horn H.M., Hansen C.J., 1988, ApJ, 325, 725

\bibitem[\protect\citeauthoryear{Messios et al.}{2001}]{Messios2001}
    Messios N., Papadopoulos D.B., Stergioulas N., 2001, MNRAS, 328, 1161 .


\bibitem[\protect\citeauthoryear{Negele \& Vautherin}{1973}]{NV1973}
   Negele J.W., Vautherin D., 1973, Nucl. Phys., A207, 298

\bibitem[\protect\citeauthoryear{Palmer et al.}{2005}]{Palmer2005}
    Palmer D.M. et al., 2005, Nature, 434, 1107

\bibitem[\protect\citeauthoryear{Pandharipande}{1971}]{EOS_A}
    Pandharipande V.R.,  1971, Nucl. Phys. A, 178, 123

\bibitem[\protect\citeauthoryear{Pandharipande \& Smith }{1975}]{EOS_L}
   Pandharipande V.R., Smith R.A., 1975, Phys. Lett. 59B, 15

\bibitem[\protect\citeauthoryear{Piro}{2005}]{Piro2005}
    Piro A.L., 2005, ApJ, 634, L153

\bibitem[\protect\citeauthoryear{Samuelsson \& Andersson}{2006}]{SA2006}
    Samuelsson L., Andersson N., astro-ph/0609265

\bibitem[\protect\citeauthoryear{Schumaker \& Thorne}{1983}]{Schumaker1983}
    Schumaker B.L., Thorne K.S., 1983, MNRAS, 203, 457

\bibitem[\protect\citeauthoryear{Strohmayer}{1991}]{Strohmayer1991a}
   Strohmayer T.E., 1991, ApJ, 372, 591

\bibitem[\protect\citeauthoryear{Sotani et al.}{2006}]{SKSV06}
   Sotani H., Kokkotas, K.D., Stergioulas, N., Vavoulidis, M., preprint (astro-ph/0611666)

\bibitem[\protect\citeauthoryear{Strohmayer et al.}{1991}]{Strohmayer1991b}
   Strohmayer T.E., etal., 1991, ApJ, 375, 679

\bibitem[\protect\citeauthoryear{Strohmayer \& Watts}{2005}]{SW2005}
   Strohmayer T.E., Watts A.L., 2005, ApJ, 632, L111

\bibitem[\protect\citeauthoryear{Strohmayer \& Watts}{2006}]{SW2006}
   Strohmayer T.E., Watts A.L., astro-ph/0608463 (accepted in ApJ)

\bibitem[\protect\citeauthoryear{Terasawa et al.}{2005}]{Terasawa2005}
    Terasawa T., et al., 2005, Nature 434, 1110

\bibitem[\protect\citeauthoryear{Unno et al.}{1989}]{Unno1989}
    Unno W., Osaki Y., Ando H., Saio H., Shibahashi H., 1989,
    Nonradial Oscillations of Stars, University of Tokyo Press

\bibitem[\protect\citeauthoryear{Wasserman \& Shapiro}{1983}]{Wasserman1983}
    Wasserman I., Shapiro S.L., 1983, ApJ, 265, 1036

\bibitem[\protect\citeauthoryear{Watts \& Strohmayer}{2006}]{WS2006}
   Watts A.L., Strohmayer T.E.,
   2006, ApJ, 637, L117

\bibitem[\protect\citeauthoryear{Wiringa et al.}{1988}]{EOS_WFF3}
    Wiringa R.B., Fiks V., Farbrocini A., 1988, Phys. Rev. C 38, 1010







\end{thebibliography}
\end{document}